\documentclass[11pt]{article}

\usepackage{color, amsmath, amssymb, amsbsy, amsthm, graphicx, bbm, amsfonts, bm, dsfont, courier, subfig}
\usepackage[toc,page]{appendix}
\usepackage{setspace}
\usepackage{booktabs}
\usepackage[flushleft]{threeparttable}
\usepackage{latexsym}
\usepackage[dvipsnames]{xcolor}
\usepackage{comment}
\usepackage{wrapfig}
\usepackage{graphicx}
\usepackage{graphics}
\usepackage{algorithm}
\usepackage{algorithmic}
\usepackage{makecell}
\usepackage[colorlinks,allcolors=blue]{hyperref}
\usepackage[round]{natbib}  
\usepackage{tikz}
\allowdisplaybreaks
\usepackage[figuresright]{rotating}
\usepackage{soul}
\usepackage{threeparttable,booktabs}
\usepackage{etoolbox}
\appto\TPTnoteSettings{\footnotesize}
\usepackage{svg} 
\usepackage{tablefootnote}
\newtheoremstyle{theoremstyle}
{\topsep} 
{\topsep} 
{\itshape} 
{} 
{} 
{} 
{.5em} 
{\color{black}\ifthenelse{\equal{#3}{}}{{\bfseries #1 #2}}{{\bfseries #1 #2 (#3)}}}
\newtheoremstyle{theoremstylealt}
{\topsep} 
{\topsep} 
{\itshape} 
{} 
{} 
{} 
{.5em} 
{\color{black}\ifthenelse{\equal{#3}{}}{{\bfseries #1 #2$^\prime$}}{{\bfseries #1 #2$^\prime$ (#3)}}}
\newtheoremstyle{examplestyle}
{\topsep} 
{\topsep} 
{} 
{} 
{} 
{} 
{.5em} 
{\color{black}\ifthenelse{\equal{#3}{}}{{\bfseries #1 #2}}{{\bfseries #1 #2 (#3)}}}
\theoremstyle{theoremstyle}\newtheorem{thm}{Theorem}
\theoremstyle{theoremstylealt}
\theoremstyle{theoremstyle}     
\theoremstyle{theoremstyle}\newtheorem{lem}{Lemma}  
\theoremstyle{theoremstyle}\newtheorem{coro}{Corollary}        
\theoremstyle{theoremstyle}
\theoremstyle{theoremstyle}\newtheorem{assumption}{Assumption}
\theoremstyle{theoremstylealt}
\theoremstyle{theoremstyle}
\theoremstyle{theoremstyle}

\theoremstyle{theoremstyle}

\theoremstyle{examplestyle}
\theoremstyle{examplestyle}
\theoremstyle{examplestyle}

\newcommand{\Expectation}{\mathbb{E}}

\newcommand{\Var}{\mathbb{V}}

\newcommand{\Indicator}{\mathbf{1}}

\newcommand{\toProb}{\overset{\mathrm{p}}{\to}}
\newcommand{\toDist}{\overset{\mathrm{d}}{\to}}
\newcommand{\precsimProb}{\precsim_{\mathrm{p}}}
\newcommand{\precProb}{\prec_{\mathrm{p}}}

\newcommand{\asympProb}{\asymp_{\mathrm{p}}}

\renewcommand{\epsilon}{\varepsilon}

\def \hat{\widehat}

\usepackage{setspace}
\usepackage{array}
\newcolumntype{H}{>{\setbox0=\hbox\bgroup}c<{\egroup}@{}}
\definecolor{applegreen}{rgb}{0.55, 0.71, 0.0}
\begin{document}

\def\spacingset#1{\renewcommand{\baselinestretch}%
{#1}\small\normalsize} \spacingset{1}
	\onehalfspacing
	\title{Mediation Analysis with Mendelian Randomization \\ and Efficient Multiple GWAS Integration
\footnotetext{Correspondence: Xinwei Ma (\href{mailto:x1ma@ucsd.edu}{x1ma@ucsd.edu}), Jingshen Wang (\href{mailto:jingshenwang@berkeley.edu}{jingshenwang@berkeley.edu}).}}

\author{Rita  Qiuran  Lyu \thanks{Division of Biostatistics, University of California Berkeley.} \and 
Chong Wu\thanks{Department of Biostatistics, The University of Texas MD Anderson Cancer Center.}  \and 
Xinwei Ma\thanks{Department of Economics, University of California San Diego.} \and
Jingshen Wang\thanks{Division of Biostatistics, University of California Berkeley.}
}

	\maketitle

	\begin{abstract}
		{
	Mediation analysis is a powerful tool for studying causal pathways between exposure, mediator, and outcome variables of interest. While classical mediation analysis using observational data often requires strong and sometimes unrealistic assumptions, such as unconfoundedness, Mendelian Randomization (MR) avoids unmeasured confounding bias by employing genetic variations as instrumental variables. We develop a novel MR framework for mediation analysis with genome-wide associate study (GWAS) summary data, and provide solid statistical guarantees. Our framework employs carefully crafted estimating equations, allowing for different sets of genetic variations to instrument the exposure and the mediator, to efficiently integrate information stored in three independent GWAS. As part of this endeavor, we demonstrate that in mediation analysis, the challenge raised by instrument selection goes beyond the well-known winner's curse issue, and therefore, addressing it requires special treatment.  
  We then develop bias correction techniques to address the instrument selection issue and commonly encountered measurement error bias issue. Collectively, through our theoretical investigations, we show that our framework provides valid statistical inference for both direct and mediation effects with enhanced statistical efficiency compared to existing methods. We further illustrate the finite-sample performance of our approach through simulation experiments and a case study. 
		}
		
	\end{abstract}
 	 \clearpage
  \onehalfspacing

\section{Introduction}\label{Section-introduction}

\subsection{Background and contribution}

Mediation analysis is a powerful tool for studying causal pathways between exposure, mediator, and outcome variables of interest. It offers a systematic framework to disentangle the direct effect of the exposure on the outcome and the indirect effect that runs through the mediator. Nevertheless, classical mediation analysis using observational data often requires strong and sometimes unrealistic assumptions, such as no unmeasured confounders in the mediator-outcome relationship. When such assumptions are violated, classical mediation analysis may produce biased causal effect estimates. While Mendelian Randomization (MR) offers a strategy to mitigate such unmeasured confounding bias by employing genetic variations as instrumental variables (IVs) \citep{carter2021mendelian}, it has been traditionally adopted to estimate the total causal effect between the exposure and outcome variables, implying a need for careful customization of MR for mediation analysis.

In this manuscript, we develop a unified Mendelian randomization framework for mediation analysis with genome-wide associate study (GWAS) summary data. We choose the research design with GWAS summary data because of the public availability of such data and the extensive breadth of traits covered by GWAS with substantial sample sizes. For example, the GWAS Catalog contains summary statistics from more than $45,000$ individual GWAS across over $6,000$ publications \citep{sollis2023nhgri}. Such resources allow researchers to analyze a broad range of exposures, mediators, and outcomes with high statistical precision, even for relatively rare outcomes. To the best of our knowledge, this is the first MR framework incorporating GWAS summary data which (i) is specifically tailored for mediation analysis, (ii) effectively addresses both the winner's and loser's curse as well as measurement error bias (detailed discussions on relevant bias sources can be found in Section \ref{sec2:set-up}), (iii) remains valid even when instrument selection is imperfect, and (iv) enjoys solid theoretical guarantees. In what follows, we discuss our contributions in more detail.

To cast some statistical insights into the potential pitfalls of mediation analysis with MR and GWAS summary data, we demonstrate the existence of two lesser-known bias issues unique to this framework. First, we show that the concept of IV selection bias in mediation analysis with MR and summary data is substantially different from what has been previously documented in the classical two-sample MR literature \citep{Ma2023BreakingTW}. In particular, this is because IV selection in mediation analysis gives rise to both ``winner's and loser's curses" (Section \ref{sec2:set-up}). While the winner's curse is more widely documented in the existing literature, the loser's curse bias, specific to mediation analysis, has never been discussed. The presence of the loser's curse bias is induced by 
the use of different genetic instruments for the mediator and the exposure variable. Second, we also demonstrate the impact of imperfect IV selection for mediation analysis using MR with summary data. While imperfect IV selection may not have a substantial impact on parameter estimation in two-sample MR, it can result in either efficiency loss or/and estimation bias in mediation analysis using MR (Section \ref{sec2:set-up}).

On the statistical methodology side, our proposed approach is not only free of the aforementioned potential sources of bias, but also efficiently integrates information stored in three independent GWAS summary data, leading to improved statistical efficiency (Theorem \ref{thm:asy efficiency gain}). As our framework successfully addresses the aforementioned bias issues, it provides valid statistical inference for both direct and mediation effects. Built upon our proposed bias correction techniques, the efficiency gain of our estimator stems from three carefully crafted estimating equations, Eq~\eqref{eq:indirect effect est equation}--\eqref{eq:direct effect est equation2}. These unique estimating equations allow us to form efficient direct effects and mediation effect estimators using solely the IVs that are either relevant to the exposure variable or the mediator. This approach differs from multivariable MR analysis, which uses the same set of genetic instruments for the exposure and the mediator. 

From a theoretical perspective, we first establish a joint asymptotic normality result for the causal effect estimators of both direct and indirect effects, allowing both the sample size and the number of instruments to diverge, as indicated in Theorem \ref{thm: variance consistency}. As these estimators are correctly centered at their population targets, this result demonstrates that our estimators are free of winner's curse bias, loser's curse bias, and measurement error bias. In addition, they remain valid even in the presence of imperfect IV selection. We next provide a consistent estimator of the covariance matrix in Theorem \ref{thm: variance consistency}, enabling us to construct valid confidence intervals for desired causal effects, as well as their combinations based on the delta method. Furthermore, we demonstrate in several important scenarios that our proposed estimators indeed offer asymptotic efficiency gain over those derived from multivariable MR analysis, as shown in Theorem \ref{thm:asy efficiency gain}.

On the practical side, we showcase the finite-sample performance of our framework through Monte Carlo experiments (Section \ref{Sec:simulation-study}) and a case study (Section \ref{Sec:real-data-analysis}). Through these results, we demonstrate that our approach (i) provides accurate estimates for both direct and mediation effects, (ii) exhibits superior performance in terms of boosted power, improved coverage, and reduced bias compared to several existing methods, and (iii) has lower variance than the (debiased) multivariable MR estimator. In the case study, our proposed approach identifies more significant pathways than the mediation analysis conducted using classical MR. 

\subsection{Existing literature}

Two MR methods using summary data are available in the literature for mediation analysis: two-step MR and multivariable MR (MVMR). However, it is unclear if these two approaches provide valid statistical inference on the direct and mediation effects. In what follows, we give a review of both methods, followed by their potential limitations. 

Two-step MR employs univariate MR analysis in a two-step fashion to estimate the mediation effect \citep{evans2015mendelian, networkmr, twostepMR}. More concretely, it first estimates the causal effect of the exposure to the mediator using the IVs that are relevant to the exposure variable, and then, similarly, estimates the causal effect of the mediator to the outcome using the  IVs that are relevant to the mediator  \citep{carter2021mendelian}. The product of the estimated effects steps gives the final mediation effect estimation. The standard errors of the mediation effect is subsequently obtained through the delta method. Note that an extra third step, a univariate MR regression of the exposure on the outcome, is needed if one is interested in decomposing the total effect of the exposure.

MVMR extends the framework of univariate MR, enabling the estimation of direct effects from multiple exposures on an outcome variable \citep{mvmr,mvmrpleiotropy,sanderson2020multivariable}. To use MVMR for estimating mediation effects, it is possible to first obtain 
 the total effect of the exposure on the outcome variable via univariate MR, and then subtract the direct effect obtained via MVMR from the estimated total effect, yielding the estimate of the mediation effect.

The above mediation analysis methods built upon two-step MR and MVMR suffer from the well-recognized measurement error and winner's curse bias issues \citep{RN16, sadreev2021navigating,Smith1981NonaggressiveBB, selectionbias}. On the one hand, the measurement error bias arises because GWAS summary statistics (i.e., associations with the instruments) are estimated with errors. Existing literature on univariate MR also refers to this measurement error bias as the weak IV bias; see \cite{sadreev2021navigating, weakIVmeasurement, Bowden159442, jiang2020measurement,zhao2020statistical, divw}, and \cite{Ma2023BreakingTW} for more discussions on how measurement error bias leads to attenuated causal effects. Unlike univariate MR, measurement error bias in MVMR and two-step MR is jointly determined by the directions and magnitudes of direct effects, making its impact on causal effect estimates complex and unclear. On the other hand,  the winner's curse arises in MR analysis whenever the same GWAS sample is used to select relevant IVs and to estimate their associations (with the exposure or the mediator). While the impact of winner's bias is well understood in univariate MR \citep{sadreev2021navigating}, its manifestation in mediation analysis using multiple GWAS is less clear.

Furthermore, solid theoretical guarantee for two-step MR and MVMR is lacking in the literature. For example, two-step MR relies on the delta method to construct standard error for the estimated mediation effect, and common practice precludes shared instruments from being used for both the exposure and the mediator. Whenever this condition fails, causal effect estimates (more precisely, estimates of $\tau_X$ and $\tau_Y$ in Figure \ref{fig:diagram}) from two-step MR can be correlated, and constructing standard errors based on two-step MR becomes problematic and challenging. A similar issue arises in MVMR, as no method is available for statistical inference on mediation effects. 

In this manuscript, we establish a unified framework for mediation analysis with multiple GWAS, in which we provide accurate direct and mediation effect estimates with valid statistical inference. 

\section{Method}\label{Section: Method}
 
\subsection{Framework overview of mediation analysis with MR and GWAS}

\begin{figure}[!ht]
		\centering
		 \begin{tikzpicture}
\node at (1, 0)  {$G_j$};
\draw (1, 0) circle (0.5);

\node at (4, 0)  {$X$};
\draw (4, 0) circle (0.5);

\node at (7, 0)  {$M$};
\draw (7, 0) circle (0.5);

\node at (10, 0)  {$Y$};
\draw (10, 0) circle (0.5);

\node at (9, 3)  {$U$};
\draw (9, 3) circle (0.5);

\draw[->] (1.65, 0)  -- (3.35, 0);
\draw[->] (4.65, 0)  -- (6.35, 0);
\draw[->] (7.65, 0)  -- (9.35, 0);

\draw[dashed, ->] (8.7, 2.4)  -- (7.3, 0.6);
\draw[dashed, ->] (9.4, 2.4)  to [bend left=15] (10, 0.7);
\draw[dashed, ->] (8.3, 3)  to [bend right=20] (4.1, 0.6);

\draw[->] (1.4, 0.5)  to [bend left=30] (6.6, 0.5);
\draw[->] (4.4, 0.5)  to [bend left=30] (9.6, 0.5);
\draw[->] (1.4, -0.5)  to [bend right=20] (9.6, -0.5);

\node at (2.8, 1.4)  {$\delta_j$};
\node at (6.5, 1.5)  {$\theta$};
\node at (5.5, -1.1)  {$\alpha_j$};
\node at (5.5, 0.2)  {$\tau_X$};
\node at (8.5, 0.2)  {$\tau_Y$};

\draw[->] (0.5, -2.5)  -- (7, -2.5);
\node at (4, -2.15)  {${\beta}_{M_j} = \tau_X{\beta}_{X_j}+\delta_j$};
\draw[->] (0.5, -3.5)  -- (10, -3.5);
\node at (5.5, -3.15)  {${\beta}_{Y_j} = \theta{\beta}_{X_j}+\tau_Y{\beta}_{M_j}+\alpha_j$};
\end{tikzpicture}
		\caption{Causal diagram for mediation analysis with multiple GWAS. }
		\label{fig:diagram}
	\end{figure}
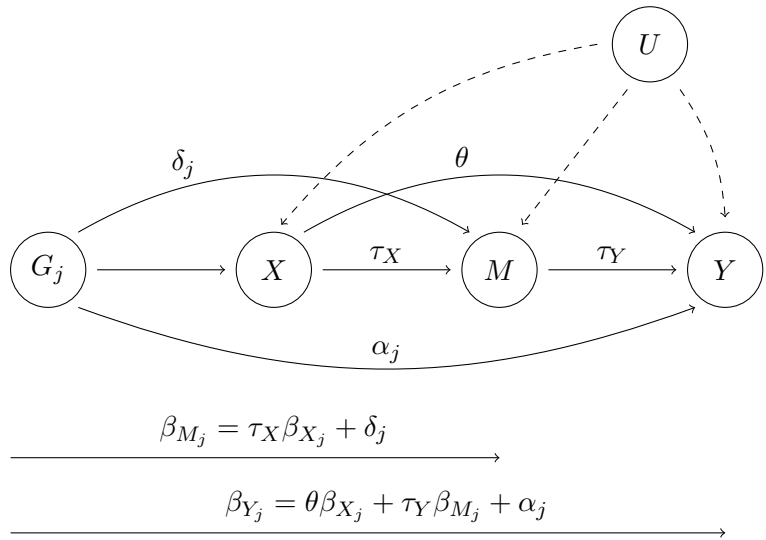

Following the causal diagram in Figure \ref{fig:diagram}, mediation analysis studies 
if the effect of an exposure $X$ on an outcome $Y$ is affected by an intermediate variable $M$, known as a mediator. This can be achieved by estimating the direct effect (denoted as $\theta$) from $X$ to $Y$, and the mediation effect (denoted as $\tau = \tau_X \tau_Y$) through the path $X\rightarrow M \rightarrow Y$. Even with individual-level data, however, a naive regression of $Y$ on the exposure $X$ or the mediator $M$ typically leads to biased causal estimates due to the presence of unmeasured confounders. Mendelian randomization (MR) avoids this issue by working with a causal pathway without the influence of $U$ through the use of genetic instruments. To be more precise, MR estimates the direct and indirect effects by leveraging genetic variations as exogenous instrument variables (IVs). Since the most common type of genetic variations among people is single nucleotide polymorphisms (SNPs), we will use the three terms, genetic variations, IVs, and SNPs, interchangeably in this manuscript. Let $G_j$ denote the $j$\textsuperscript{th} SNP, and its associations with the exposure, the mediator, and the outcome variable as $\beta_{X_j}$, $\beta_{M_j}$, and $\beta_{Y_j}$, respectively. Then, at the population level, MR works with the following structure equations:
  \begin{align}
	{\beta}_{Y_j} =&\label{eq:structural-eq-oracle1} \beta_{X_j} \cdot \theta\cdot\Indicator_{(j\in  \mathcal{S}_{\mathtt{x}}^*)} + \beta_{M_j}\cdot \tau_Y \cdot\Indicator_{(j\in  \mathcal{S}_{\mathtt{m}}^*)}+  \alpha_j,\\
 {\beta}_{M_j} =& \label{eq:structural-eq-oracle2}\beta_{X_j} \cdot \tau_X \cdot\Indicator_{(j\in  \mathcal{S}_{\mathtt{x}}^*)} +\delta_j.
	\end{align}  
In above equations,  $\mathcal{S}_{\mathtt{x}}^*$ and $\mathcal{S}_{\mathtt{m}}^*$ are indices for the relevant IV of the exposure $X$ and mediator $M$, which are formally defined as
\begin{align*}
    \mathcal{S}_{\mathtt{x}}^*=\left\{j:\beta_{X_j}\neq 0, \ j=1, \cdots, p\right\}, \quad \mathcal{S}_{\mathtt{m}}^*=\left\{j:\beta_{M_j}\neq 0, \ j=1, \cdots, p\right\}. 
\end{align*} 

In this manuscript, we consider a summary-data setting for conducting mediation analysis using MR, which employs information stored in genome-wide association studies (GWAS). In GWAS, sample analogs of $\beta_{X_j}$, $\beta_{M_j}$,  $\beta_{Y_j}$ are obtained by regressing the exposure, outcome, and mediator separately on each SNP in the corresponding datasets and then recording marginal regression coefficients and standard errors. We impose the following high-level assumption on the GWAS summary data, which is in line with the existing literature \cite{zhao2020statistical, divw}:

\begin{assumption}  [\normalfont Estimated association effects]
		\label{Assumption-1: Estimated association effects}
The estimated association effects are mutually independent and follow normal distributions:
\begin{align*}
\hat\beta_{X_j} \sim \mathcal{N}(\beta_{X_j}, \sigma_{X_j}^2),\quad \hat\beta_{M_j} \sim \mathcal{N}(\beta_{M_j}, \sigma_{M_j}^2),\quad \hat\beta_{Y_j} \sim \mathcal{N}(\beta_{Y_j}, \sigma_{Y_j}^2).
\end{align*}
For any SNP $j'\neq j$, $(\hat{\beta}_{X_j},\hat{\beta}_{M_j},\hat{\beta}_{Y_j})$ and $(\hat{\beta}_{X_{j'}},\hat{\beta}_{M_{j'}},\hat{\beta}_{Y_{j'}})$ are independent. 
In addition, there exists some $\nu\to 0$, such that $\{\sigma_{Y_j}/\nu,\ \sigma_{X_j}/\nu,\ \sigma_{M_j}/\nu:1\leq j\leq p\}$ are bounded and bounded away from zero.\end{assumption}

We now briefly discuss this assumption. First, it requires that the estimated SNP-exposure, SNP-mediator, and SNP-outcome associations are statistically independent. This is typically guaranteed by employing three different GWAS for harvesting those association estimates. For future reference, we will refer the three GWAS as GWAS (I), (II), and (III), containing $\{\hat{\beta}_{X_j}:j=1,2,\dots,p\}$, $\{\hat{\beta}_{Y_j}:j=1,2,\dots,p\}$, and $\{\hat{\beta}_{M_j}:j=1,2,\dots,p\}$, respectively. Second, the assumption also imposes independence across different SNPs, which is typically guaranteed by employing SNPs that are not in linkage disequilibrium. Finally, the normal distribution assumption stems from the fact that association estimates are obtained with large samples. We also make the additional assumption that standard errors of the association estimates are of similar magnitude, which greatly simplifies the theoretical discussion. 

To estimate the parameters of interest, existing literature directly replaces the structure equation Eq \eqref{eq:structural-eq-oracle1}--\eqref{eq:structural-eq-oracle2} with the sample analog \citep{sanderson2020multivariable, carter2021mendelian}. For example, applying such a direct ``plug-in'' approach to \eqref{eq:structural-eq-oracle2} leads to an inverse variance weighting (IVW) estimator for $\tau_X$ under the classical two-sample MR framework. However, in the following section, we will discuss the unique challenges imposed by such practices for mediation analysis. 

\subsection{Unique challenges raised by unknown $\mathcal{S}_{\mathtt{x}}^*$ and $\mathcal{S}_{\mathtt{m}}^*$}\label{sec2:set-up}

In this section, we emphasize two unique challenges encountered in mediation analysis with MR and GWAS summary data raised by ``imperfect IV selection" and ``loser's curse bias." Briefly, if both  $\mathcal{S}_{\mathtt{x}}^*$ and $\mathcal{S}_{\mathtt{m}}^*$ are known a priori, it is natural to replace the population associations with the measured associations from the three GWAS. It is then possible to estimate $\theta$ and $\tau_Y$ by regressing $\hat{\beta}_{Y_j}$ on $\hat{\beta}_{X_j}\Indicator_{(j\in  \mathcal{S}_{\mathtt{x}}^*)}$ and $\hat{\beta}_{M_j}\Indicator_{(j\in  \mathcal{S}_{\mathtt{m}}^*)}$, and estimate $\tau_X$ by regressing $\hat{\beta}_{M_j}$ on $\hat{\beta}_{X_j}$ for $j \in \mathcal{S}_{\mathtt{x}}^*$. Unfortunately, not only are the true indices $\mathcal{S}_{\mathtt{x}}^*$ and $\mathcal{S}_{\mathtt{m}}^*$ unknown, but using their empirical estimates can also introduce efficiency loss or omitted IV bias due to imperfect selection, winner's and loser's curse biases, and measurement error bias. 

To our knowledge, the imperfect IV selection issue has never been recognized in the existing mediation analysis literature. Importantly, this is unique to mediation analysis because relevant genetic variations for the exposure are usually different from that for the mediator (i.e., $\mathcal{S}_{\mathtt{x}}^*\neq \mathcal{S}_{\mathtt{m}}^*$). For the under-selection case, the bias can arise whenever the selection procedure omits some relevant IVs in $\mathcal{S}_{\mathtt{x}}^*$ or $\mathcal{S}_{\mathtt{m}}^*$. Please see our Supplementary Material (Section 5) for the derivation of this claim, where we have shown the simple plug-in estimates of the causal effects using Eq~\eqref{eq:structural-eq-oracle1} are biased unless IV selection is perfect. For the over-selection case, including irrelevant IVs results in estimating $\theta$ (or $\tau_Y$) using IVs that are unrelated to the exposure (or the mediator). Both under- and over-selection lead to either bias or a loss of statistical efficiency in estimating causal effects. We note that the classical two-sample MR for estimating a single causal effect does not suffer from the imperfect IV selection issue. 

Next, different from the two-sample MR for estimating a single causal effect, mediation analysis with MR suffers from both winner's and loser's curse biases. The widely discussed winner's curse bias occurs because we select SNPs based on their standardized effect sizes stored in the GWAS summary data, and subsequently use the same summary data to estimate the causal effect. The classical two-sample MR literature documents that this data double usage often pushes the estimated effects of selected SNPs away from the true associations \citep{selectionbias, RN16}. In statistical terms, take a SNP $j$ and its estimated association with the exposure $\hat{\beta}_{X_j}$ as an example. This SNP is selected to estimate the downstream causal effect only when its standardized association effect size, \(| \hat{\beta}_{X_j}/\sigma_{X_j} |\), surpasses a given threshold \(\lambda\). The typical winner's curse in two-sample MR refers to \(\mathbb{E}[\hat{\beta}_{X_j} \mid | \hat{\beta}_{X_j}/\sigma_{X_j} | > \lambda] \neq \beta_{X_j}\), meaning that the estimated association no longer accurately centers around \(\beta_{X_j}\) after conditioning on this SNP being selected.

The loser's curse is much less documented. We can understand its presence from the structural equation $ {\beta}_{M_j} =\beta_{X_j} \cdot \tau_X \cdot \Indicator_{(j\in  \mathcal{S}_{\mathtt{x}}^*)} +\delta_j$. Implementing this equation requires one to first select relevant genetic instruments for the exposure $X$ (for example, through  thresholding $|\hat{\beta}_{X_j}/\sigma_{X_j} | > \lambda$) and then employ those selected instruments to form an inverse variance weighting type estimator. Among those genetic instruments, however, some may not be relevant for the mediator. This does not pose any issue in standard two-sample MR analysis, as there is no need to explicitly select instruments for the mediator if the only parameter of interest is the direct effect $\tau_X$. In mediation analysis, on the other hand, it is necessary to select instruments for both the exposure and the mediator to accurately estimate the other two direct effects $\theta$ and $\tau_Y$, leading to the loser's curse: $\Expectation\big[\hat{\beta}_{M_j}\big| | \hat{\beta}_{M_j}/\sigma_{M_j} | <\lambda \big]\neq \beta_{M_j}$. Our proposed method corrects for both winner's and loser's curse biases, and hence it can accommodate different genetic variations being used to instrument the exposure and the mediator. To compare, the multivariable MR approach requires a common set of IVs for $X$ and $M$. 

Lastly, the measurement error bias, which is also known as the weak IV bias in the literature \citep{sadreev2021navigating, weakIVmeasurement, Bowden159442, jiang2020measurement}, occurs because  $\Expectation[\hat{\beta}_{X_j}^2] = \beta_{X_j}^2+\sigma_{X_j}^2 > \beta_{X_j}^2$. The measurement error bias issue will be further complicated by instrument selection. In particular, as we will show in the next subsection, incorporating IV selection and correcting for this bias will call for unbiased estimators for $\beta_{X_j}^2$ and $\beta_{M_j}^2$ conditioning on the full selection events. 

\subsection{Proposed method: MAGIC}\label{Sec3:Proposed-framework}

In this subsection, we introduce our \underline{M}ediation \underline{A}nalysis framework through \underline{G}WAS summary data \underline{I}ntegration with bias \underline{C}orrection, thus with an acronym MAGIC. 

As we have discussed, a naive implementation of \eqref{eq:structural-eq-oracle1} and \eqref{eq:structural-eq-oracle2} may lead to biased causal estimates due to imperfect IV selection. Recall this bias is caused by \eqref{eq:structural-eq-oracle1} involving two potentially different sets of instruments, one for the exposure ($\mathcal{S}_{\mathtt{x}}^*$) and the other for the mediator ($\mathcal{S}_{\mathtt{m}}^*$). The errors in selecting these two sets will interact in complicated ways, leading to the imperfect IV selection bias (see Section 5 in the Supplementary Material for rigorous derivations). We tackle the imperfect IV selection issue by constructing three estimating equations:
\begin{alignat}{2}
\label{eq:indirect effect est equation}       &\sum\limits_{j\in  \mathcal{S}_{\mathtt{x}}^*}\beta_{X_j}({\beta}_{M_j}-\delta_j) =\sum\limits_{j\in  \mathcal{S}_{\mathtt{x}}^*} \beta_{X_j}^2\cdot \tau_X, &&\\
\label{eq:direct effect est equation}    &\sum\limits_{j\in  \mathcal{S}_{\mathtt{x}}^*}\beta_{X_j}({\beta}_{Y_j}-\alpha_j)= \sum\limits_{j\in  \mathcal{S}_{\mathtt{x}}^*}\beta_{X_j}^2 \cdot \theta  &&+ \sum\limits_{j\in  \mathcal{S}_{\mathtt{x}}^*}\beta_{X_j}\beta_{M_j}\cdot \tau_Y ,\\ 
\label{eq:direct effect est equation2} & \sum\limits_{j\in  \mathcal{S}_{\mathtt{m}}^*}\beta_{M_j}({\beta}_{Y_j}-\alpha_j)= \sum\limits_{j\in  \mathcal{S}_{\mathtt{m}}^*}\beta_{X_j}\beta_{M_j}\cdot \theta &&+ \sum\limits_{j\in  \mathcal{S}_{\mathtt{m}}^*}\beta_{M_j}^2 \cdot \tau_Y .
\end{alignat}  
The first estimating equation in Eq \eqref{eq:indirect effect est equation} is a straightforward re-arrangement of \eqref{eq:structural-eq-oracle2}. The second and the third estimating equations in (i.e., Eq \eqref{eq:direct effect est equation} and \eqref{eq:direct effect est equation2}) are derived from \eqref{eq:structural-eq-oracle1}. Importantly, as each estimating equation we propose above depends on only one set of instruments, once the unknown association effects are replaced by unbiased sample analogues, they produce accurate causal effect estimates even in the presence of imperfectly selected IVs. At the same time, we still allow the three estimating equations to employ different sets of IVs which, as we show later, can lead to efficiency gains. 

Furthermore, as shall be made clear in our theoretical investigation and simulation studies, when the relevant IV indices are known, our proposed estimator has the same performance as the one derived from solving the sample analog of Eq \eqref{eq:structural-eq-oracle1}--\eqref{eq:structural-eq-oracle2}. When the estimated indices omit relevant IVs or include irrelevant IVs, while the performance of our estimator is unaffected, the estimator derived from solving the sample analog of Eq \eqref{eq:structural-eq-oracle1}--\eqref{eq:structural-eq-oracle2} is heavily biased. This is borne out in both simulation studies and our theoretical investigation. See the Supplementary Material for details. 

To operationalize the estimating equations in \eqref{eq:indirect effect est equation}--\eqref{eq:direct effect est equation2}, naively replacing the population association effects $\beta_{X_j}$, $\beta_{M_j}$ and $\beta_{Y_j}$ with their empirical estimates from GWAS will not provide causal effect estimates due to the presence of the loser's curse, winner's curse and measurement error bias. In what follows, we propose a strategy that systematically removes all three biases. 

First, we shall propose a new bias-corrected estimator, \(\hat{\beta}_{X_j,\mathtt{bc}}\) and \(\hat{\beta}_{M_j,\mathtt{bc}}\), designed to eliminate both the loser's and winner's curse. These estimators are constructed to ensure that:
\[
\mathbb{E}[\hat{\beta}_{X_j,\mathtt{bc}} \mid {\mathcal{S}}_{\mathtt{x}}, {\mathcal{S}}_{\mathtt{m}}] = \beta_{X_j} \quad \text{and} \quad \mathbb{E}[\hat{\beta}_{M_j,\mathtt{bc}} \mid {\mathcal{S}}_{\mathtt{x}}, {\mathcal{S}}_{\mathtt{m}}] = \beta_{M_j},
\]
where \({\mathcal{S}}_{\mathtt{x}}\) and \({\mathcal{S}}_{\mathtt{m}}\) represent the selected sets of IVs associated with \(X\) and \(M\), respectively (see Section \ref{section: bias correction} below for detailed construction). In other words, we will propose estimating equations that are unbiased when conditioning on the full selection events, providing immunity to both the winner's curse and the loser's curse. As mentioned earlier, this dual immunity is crucial for the accurate estimation of mediation effects and for ensuring valid statistical inference. 

Second, as the proposed estimating equations defined in Eq \eqref{eq:indirect effect est equation}--\eqref{eq:direct effect est equation2} involve $\beta_{X_j}^2$ and $\beta_{M_j}^2$, using their 
bias-corrected sample analogues directly will introduce measurement error bias into these estimating equations. More precisely, this is because $\Expectation[\hat{\beta}_{X_j,\mathtt{bc}}^2 | {\mathcal{S}}_{\mathtt{x}},{\mathcal{S}}_{\mathtt{m}}] = \beta_{X_j}^2 + \Var[\hat{\beta}_{X_j,\mathtt{bc}}^2 | {\mathcal{S}}_{\mathtt{x}},{\mathcal{S}}_{\mathtt{m}}] > \beta_{X_j}^2$ and $\Expectation[\hat{\beta}_{M_j,\mathtt{bc}}^2 | {\mathcal{S}}_{\mathtt{x}},{\mathcal{S}}_{\mathtt{m}}] = \beta_{M_j}^2 + \Var[\hat{\beta}_{M_j,\mathtt{bc}}^2 | {\mathcal{S}}_{\mathtt{x}},{\mathcal{S}}_{\mathtt{m}}] > \beta_{M_j}^2$. 
Although it is possible to theoretically derive the post-selection variances ($\Var[\hat{\beta}_{X_j,\mathtt{bc}}^2 | {\mathcal{S}}_{\mathtt{x}},{\mathcal{S}}_{\mathtt{m}}]$ and $ \Var[\hat{\beta}_{M_j,\mathtt{bc}}^2 | {\mathcal{S}}_{\mathtt{x}},{\mathcal{S}}_{\mathtt{m}}] $) of the bias-corrected association effect estimators, unfortunately, they take complicated forms and nonlinearly depend on the unknown associations ($\beta_{X_j}$ and $\beta_{M_j}$). Instead of directly estimating the conditional variances of $\hat{\beta}_{X_j,\mathtt{bc}}$ and $\hat{\beta}_{M_j,\mathtt{bc}}$, we construct estimates of the bias. To be precise, we provide below $\hat{\varsigma}_{X_j}$ and $\hat{\varsigma}_{M_j}$ satisfying 
\begin{align}\label{eq:square-bias correction}
\Expectation[\hat{\beta}_{X_j,\mathtt{bc}}^2 - \hat{\varsigma}_{X_j} | {\mathcal{S}}_{\mathtt{x}},{\mathcal{S}}_{\mathtt{m}}] = \beta_{X_{j}}^2,\quad \Expectation[\hat{\beta}_{M_j,\mathtt{bc}}^2 - \hat{\varsigma}_{M_j} | {\mathcal{S}}_{\mathtt{x}},{\mathcal{S}}_{\mathtt{m}}] = \beta_{M_{j}}^2. 
\end{align}

Lastly, combining the previous two bias correction techniques, our sample estimating equations for the direct effects are 
\begin{alignat*}{3}
   &\sum_{j\in\mathcal{S}_{\mathtt{x}}} \frac{\hat{\beta}_{M_j,\mathtt{bc}}\hat\beta_{X_j,\mathtt{bc}}}{\sigma_{M_j}^2}\ =\  &&\hat{\tau}_X\sum_{j\in\mathcal{S}_{\mathtt{x}}}\frac{\hat{\beta}_{X_j,\mathtt{bc}}^2 - \hat{\varsigma}_{X_j} }{\sigma_{M_j}^2} &&,\\[12pt] 
		&\sum_{j\in\mathcal{S}_{\mathtt{x}}} \frac{\hat\beta_{Y_j}\hat{\beta}_{X_j,\mathtt{bc}}}{\sigma_{Y_j}^2}\ =\  &&\hat{\theta}\sum_{j\in\mathcal{S}_{\mathtt{x}}}\frac{\hat{\beta}_{X_j,\mathtt{bc}}^2 - \hat{\varsigma}_{X_j}}{\sigma_{Y_j}^2}&&\ +\  \hat{\tau}_{Y}\sum_{j\in\mathcal{S}_{\mathtt{x}}}\frac{\hat{\beta}_{X_j,\mathtt{bc}}\hat{\beta}_{M_j,\mathtt{bc}} }{\sigma_{Y_j}^2}, \\[12pt]
		&\sum_{j\in\mathcal{S}_{\mathtt{m}}} \frac{\hat\beta_{Y_j}\hat{\beta}_{M_j,\mathtt{bc}}}{\sigma_{Y_j}^2}\ =\  &&\hat{\theta}\quad \sum_{j\in\mathcal{S}_{\mathtt{m}}}\frac{\hat{\beta}_{X_j,\mathtt{bc}}\hat{\beta}_{M_j,\mathtt{bc}}}{\sigma_{Y_j}^2} &&\ +\  \hat{\tau}_{Y}\sum_{j\in\mathcal{S}_{\mathtt{m}}}\frac{\hat{\beta}_{M_j,\mathtt{bc}}^2 - \hat{\varsigma}_{M_j} }{\sigma_{Y_j}^2}.
\end{alignat*}
Here, we also incorporate normalization by $\sigma_{M_j}^2$ and $\sigma_{Y_j}^2$, following current practice of Mendelian randomization. The estimated direct effects can also be written in matrix form:
\begin{align*}
 \left[\begin{array}{c}
		\widehat{\theta} \\
		\widehat{\tau}_{Y}\\
  \widehat{\tau}_{X}
	\end{array}\right]= \underbrace{\left[\begin{array}{ccc}
		\sum\limits_{j \in \mathcal{S}_{\mathtt{x}}}\frac{\widehat{\beta}_{X_j, \mathtt{bc}}^2-\hat{\varsigma}_{X_j} }{\sigma_{Y_j}^2} & \sum\limits_{j \in \mathcal{S}_{\mathtt{x}}} \frac{\widehat{\beta}_{X_j, \mathtt{bc}} \widehat{\beta}_{M_j, \mathtt{bc}} }{\sigma_{Y_j}^2}&0 \\[12pt]
		\sum\limits_{j \in \mathcal{S}_{\mathtt{m}}} \frac{\widehat{\beta}_{X_j, \mathtt{bc}} \widehat{\beta}_{M_j, \mathtt{bc}} }{\sigma_{Y_j}^2} & \sum\limits_{j \in \mathcal{S}_{\mathtt{m}}}\frac{\widehat{\beta}_{M_j, \mathtt{bc}}^2-\hat{\varsigma}_{M_j} }{\sigma_{Y_j}^2}&0\\[12pt]
  0&0&\sum\limits_{j \in \mathcal{S}_{\mathtt{x}}}\frac{\widehat{\beta}_{X_j, \mathtt{bc}}^2-\hat{\varsigma}_{X_j} }{\sigma_{M_j}^2} 
	\end{array}\right]^{-1}}_{\textstyle=: \hat{\mathsf{M}}^{-1} }\left[\begin{array}{c}
		\sum\limits_{j \in \mathcal{S}_{\mathtt{x}}} \frac{\widehat{\beta}_{Y_j} \widehat{\beta}_{X_j, \mathtt{bc}}} { \sigma_{Y_j}^2 }\\[12pt]
		\sum\limits_{j \in \mathcal{S}_{\mathtt{m}}} \frac{\widehat{\beta}_{Y_j} \widehat{\beta}_{M_j, \mathtt{bc}} }{\sigma_{Y_j}^2}\\[12pt]
  \sum\limits_{j \in \mathcal{S}_{\mathtt{x}}} \frac{ \widehat{\beta}_{M_j, \mathtt{bc}} \widehat{\beta}_{X_j, \mathtt{bc}} }{ \sigma_{M_j}^2}
	\end{array}\right].
\end{align*}
The mediation effect, $\tau$, can be estimated by the product 
\begin{align*}
    \hat{\tau}=\hat{\tau}_{X}\hat{\tau}_{Y}.
\end{align*}

As we discussed in Introduction, another important contribution of this manuscript is that we provide a joint asymptotic normality characterization of the estimated direct effects, together with an estimator for the asymptotic covariance matrix. Collectively, they lead to valid statistical inference on not only the direct effects but also transformations therefore, such as the mediation and total effects via the delta method. Therefore, we provide a suite of tools for mediation analysis, including estimation and statistical inference, in the summary-data Mendelian randomization framework. In what follows, we first provide the covariance estimator, and then discuss in more detail our bias correction technique. 

Our covariance estimator builds on the idea of ``regression residuals.'' We provide below its precise construction, while a formal theoretical justification is postponed to the next section. We first define  
\begin{align*}
    \hat{U}_j = \begin{bmatrix}
          \hat{u}_{j,\theta} \\
          \hat{u}_{j,\tau_Y}\\
          \hat{u}_{j,\tau_X}
    \end{bmatrix} = \begin{bmatrix}
        \Indicator_{(j\in \mathcal{S}_{\mathtt{x}})}\cdot\frac{\widehat{\beta}_{X_j, \mathtt{bc}}\big(\hat{\beta}_{Y_j}  -\hat{\tau}_Y \hat{\beta}_{M_j, \mathtt{bc}}\big)+\hat{\theta} \big(\hat{\varsigma}_{X_j}  - \widehat{\beta}_{X_j, \mathtt{bc}}^2\big)}{\sigma_{Y_j}^2}\\[12pt]
        \Indicator_{(j\in \mathcal{S}_{\mathtt{m}})}\cdot\frac{\widehat{\beta}_{M_j, \mathtt{bc}}\big(\hat{\beta}_{Y_j}  -\hat{\theta} \hat{\beta}_{X_j, \mathtt{bc}}\big)+\hat{\tau}_Y \big(\hat{\varsigma}_{M_j} - \widehat{\beta}_{M_j, \mathtt{bc}}^2\big)}{\sigma_{Y_j}^2}\\[12pt]
         \Indicator_{(j\in \mathcal{S}_{\mathtt{x}})}\cdot\frac{\hat{\beta}_{X_j, \mathtt{bc}}\widehat{\beta}_{M_j, \mathtt{bc}}+\hat{\tau}_X \big(\hat{\varsigma}_{X_j} - \widehat{\beta}_{X_j, \mathtt{bc}}^2\big)}{\sigma_{M_j}^2} 
    \end{bmatrix} .
\end{align*}
Then, the covariance estimator takes the form
\begin{align*}
    \hat{\mathsf{V}} &=  \hat{\mathsf{M}}^{-1} \hat{\mathsf{U}}(\hat{\mathsf{M}}^{-1})^\intercal,
\end{align*}
where the middle matrix is $\hat{\mathsf{U}} = \sum_{j=1}^p \hat{U}_j\hat{U}_j^\intercal$. Asymptotic variance of the estimated mediation effect, $\hat{\tau}$, can be obtained via the delta method:
\begin{align*}
\hat{\mathsf{v}}_{\tau}=\hat{\mathsf{c}}^{\intercal}\hat{\mathsf{V}}\hat{\mathsf{c}}, \text{ where } \hat{\mathsf{c}}=(0,\hat\tau_{X},\hat\tau_{Y})^{\intercal}. 
\end{align*}

\subsection{Bias corrected association effect estimation}\label{section: bias correction}

In the remainder of this section, we provide the specific forms of the bias-corrected association effect estimators $\widehat{\beta}_{M_j, \mathtt{bc}}$ and $\widehat{\beta}_{X_j, \mathtt{bc}}$, as well as our instrument selection method. To start, define the pseudo SNP-exposure associations $Z_1, \dots, Z_p$ and the pseudo SNP-mediator associations $Z_1', \dots, Z_p'$, following the $\mathcal{N}(0, \eta^2)$ distribution. Here, $\eta$ is a pre-specified positive constant. With fixed cutoff value $\lambda > 0$, the selected instruments are:
\begin{align}\label{eq:selection}
    &\mathcal{S}_{\mathtt{x}}=\Big\{j:\Big| \frac{\hat{\beta}_{X_j}}{ \sigma_{X_j} } + Z_j \Big| > \lambda, \ j=1, \ldots, p\Big\} \text{ and } \mathcal{S}_{\mathtt{m}}=\Big\{j:\Big| \frac{\hat{\beta}_{M_j}}{ \sigma_{M_j} } + Z'_j \Big| > \lambda, \ j=1, \ldots, p\Big\}.
\end{align} 
To define the bias-corrected association effects estimators, let $ A_{j,\pm} =  -  \frac{\hat{\beta}_{X_j}}{ \sigma_{X_j}\eta } \pm\frac{\lambda}{\eta}$ and $A_{j,\pm}' =  -  \frac{\hat{\beta}_{M_j}}{ \sigma_{M_j}\eta } \pm\frac{\lambda}{\eta}$. Then we propose
\begin{align*}
\hat{\beta}_{X_j,\mathtt{bc}} =&\  \hat{\beta}_{X_j} - \frac{\sigma_{X_j}}{\eta}\Big(\phi\big(A_{j,+}\big) - \phi\big(A_{j,-}\big)\Big)\Big(\frac{\Indicator_{(j\in\mathcal{S}_{\mathtt{x}})}}{1 - \Phi\big(A_{j,+}\big) + \Phi\big(A_{j,-}\big)} - \frac{\Indicator_{(j\not\in\mathcal{S}_{\mathtt{x}})}}{ \Phi\big(A_{j,+}\big) - \Phi\big(A_{j,-}\big)} \Big), \\
\hat{\varsigma}_{X_j}  =&\ \sigma_{X_j}^2\Bigg( 1 - \frac{1}{\eta^2}  \frac{A_{j,+}\phi(A_{j,+}) - A_{j,-}\phi(A_{j,-})}{1 - \Phi(A_{j,+}) + \Phi( A_{j,-} )} +\frac{1}{\eta^2} \Big(\frac{\phi(A_{j,+}) - \phi(A_{j,-}) }{1 - \Phi(A_{j,+}) + \Phi( A_{j, -} )}\Big)^2 \Bigg)\Indicator_{(j\in\mathcal{S}_{\mathtt{x}})} \\
&+ \sigma_{X_j}^2\Bigg( 1 - \frac{1}{\eta^2}  \frac{{-A_{j,+}\phi(A_{j,+}) + A_{j,-}\phi(A_{j,-})}}{\Phi(A_{j,+}) - \Phi( A_{j,-} )} +\frac{1}{\eta^2} \Big(\frac{-\phi(A_{j,+}) +\phi(A_{j,-}) }{ \Phi(A_{j,+}) - \Phi( A_{j, -} )}\Big)^2 \Bigg)\Indicator_{(j\not\in\mathcal{S}_{\mathtt{x}})},
\end{align*}
and 
\begin{align*}
\hat{\beta}_{M_j,\mathtt{bc}} =& \  \hat{\beta}_{M_j} - \frac{\sigma_{M_j}}{\eta}\Big(\phi\big(A_{j,+}'\big) - \phi\big(A_{j,-}'\big)\Big)\Big(\frac{\Indicator_{(j\in\mathcal{S}_{\mathtt{m}})}}{1 - \Phi\big(A_{j,+}'\big) + \Phi\big(A_{j,-}'\big)} - \frac{\Indicator_{(j\not\in\mathcal{S}_{\mathtt{m}})}}{ \Phi\big(A_{j,+}'\big) - \Phi\big(A_{j,-}'\big)} \Big), \\
\hat{\varsigma}_{M_j}  =&\ \sigma_{M_j}^2\Bigg( 1 - \frac{1}{\eta^2}  \frac{A_{j,+}'\phi(A_{j,+}') - A_{j,-}'\phi(A_{j,-}')}{1 - \Phi(A_{j,+}') + \Phi( A_{j,-}' )} +\frac{1}{\eta^2} \Big(\frac{\phi(A_{j,+}') - \phi(A_{j,-}') }{1 - \Phi(A_{j,+}') + \Phi( A_{j, -}' )}\Big)^2 \Bigg)\Indicator_{(j\in\mathcal{S}_{\mathtt{m}})} \\
&+\sigma_{M_j}^2\Bigg( 1 - \frac{1}{\eta^2}  \frac{{{-A_{j,+}'\phi(A_{j,+}') +A_{j,-}'\phi(A_{j,-}')}}}{ \Phi(A_{j,+}') - \Phi( A_{j,-}' )} +\frac{1}{\eta^2} \Big(\frac{-\phi(A_{j,+}') + \phi(A_{j,-}') }{\Phi(A_{j,+}') - \Phi( A_{j, -}' )}\Big)^2 \Bigg)\Indicator_{(j\not\in\mathcal{S}_{\mathtt{m}})},
\end{align*}  
where $\Phi(\cdot)$ and $\phi(\cdot)$ are the cumulative distribution function and the probability density function of the standard normal distribution. It is worth mentioning that while our construction is inspired by the framework of \cite{Ma2023BreakingTW}, it tackles both the winner's and the loser's curse. To compare, they focus on the standard two-sample MR setting, and their method is only able to remove the winner's curse bias. By ensuring that the bias-corrected association effects are immune to both winner's and loser's curse biases, our proposed method allows for accurate causal effect estimation and reliable statistical inference in mediation analysis.

\section{Theoretical investigation}

In this section, we provide theoretical justifications for our proposed methodology. We first show that our bias-corrected estimators lift both the winner's curse and the loser's curse (Lemma \ref{lem: RB unbiasedness and variance validity}). We then demonstrate the validity of our statistical inference procedure by proving a joint asymptotic normality result for the estimated direct effects and by establishing  consistency of the variance-covariance matrix estimator (Theorem \ref{thm: variance consistency}).  Finally, we illustrate that our estimator is more efficient asymptotically compared to the multivariable MR approach in important settings (Theorem \ref{thm:asy efficiency gain}).  

\subsection{Notations and assumptions} \label{Sec:notations}

We start by introducing the following notation for probabilistic ordering. We consider the asymptotic regime where $p\rightarrow \infty$. For two sequences of random variables, $A$ and $B$, write $A \precsimProb B$ if the ratio $A/B$ is asymptotically bounded in probability. The strict relation, $A\precProb B$, implies that $A/B\toProb 0$. Finally, $A\asympProb B$ indicates both $A\precsimProb B$ and $B\precsimProb A$. 

Recall from the previous section that we defined two selected IV sets $\mathcal{S}_{\mathtt{x}}$ and $\mathcal{S}_{\mathtt{m}}$ in Eq~\eqref{eq:selection}, after conducting rerandomized selection from GWAS (I) and (III). Denote $p_\mathtt{x}$ and $p_\mathtt{m}$ as the cardinalities of $\mathcal{S}_\mathtt{x}$ and $\mathcal{S}_\mathtt{m}$, respectively.  For the selected genetic instruments, their average strengths are
\begin{align*}
    \kappa_{\mathtt{x}} = \frac{1}{p_{\mathtt{x}}}\sum\limits_{j\in \mathcal{S}_{\mathtt{x}}}\left(\frac{\beta_{X_j}}{\sigma_{X_j}}\right)^2 \quad \text{ and } \quad \kappa_{\mathtt{m}} = \frac{1}{p_{\mathtt{m}}}\sum\limits_{j\in \mathcal{S}_{\mathtt{m}}}\left(\frac{\beta_{M_j}}{\sigma_{M_j}}\right)^2.
\end{align*}
As our procedure explicitly allows for different sets of instruments for the exposure and the mediator, it is also helpful to define
\begin{align*}
    &\kappa^{\mathtt{m}}_{\mathtt{x}} = \frac{1}{p_{\mathtt{x}}}\sum\limits_{j\in \mathcal{S}_{\mathtt{x}}}\left(\frac{\beta_{M_j}}{\sigma_{M_j}}\right)^2,\quad
    \kappa^{\mathtt{x}}_{\mathtt{m}} = \frac{1}{p_{\mathtt{m}}}\sum\limits_{j\in \mathcal{S}_{\mathtt{m}}}\left(\frac{\beta_{X_j}}{\sigma_{X_j}}\right)^2,\quad
    \kappa^{\delta}_{\mathtt{x}} = \frac{1}{p_{\mathtt{x}}}\sum\limits_{j\in \mathcal{S}_{\mathtt{x}}}\left(\frac{\delta_j}{\sigma_{M_j}}\right)^2,\quad
\kappa^{\delta}_{\mathtt{m}} = \frac{1}{p_{\mathtt{m}}}\sum\limits_{j\in \mathcal{S}_{\mathtt{m}}}\left(\frac{\delta_j}{\sigma_{M_j}}\right)^2.
\end{align*}
While it is possible to allow the kappas to have different asymptotic order, this will necessarily lead to more cumbersome notation and lengthy theoretical discussions. To simplify the presentation, we make the following assumptions on the number of selected instruments and average instrument strength. 

\begin{assumption}[\normalfont IV selection and strength]\label{assumption:IV-strength}
	\label{Assumption-2:IV selection and strength } 
There exist $\mathsf{p}\to\infty$ and $\kappa\succsim 1$, such that (i) $p_{\mathtt{x}}, p_{\mathtt{m}}\asympProb \mathsf{p}$; (ii) $ \kappa_{\mathtt{x}},\kappa_{\mathtt{m}}, \kappa^{\delta}_{\mathtt{m}} \asympProb \kappa$, and $\kappa^{\mathtt{m}}_{\mathtt{x}}, \kappa^{\mathtt{x}}_{\mathtt{m}},\kappa^{\delta}_{\mathtt{x}} \precsimProb \kappa$. 
\end{assumption} 

In mediation analysis, pleiotropic effects play two roles. On the one hand, the pleiotropic effects on the mediator $M$ (i.e., $\delta_j$) are part of the instrument effects; that is, they help disentangle the direct effect of the mediator on the outcome ($\tau_Y$) from the direct effect of the exposure ($\theta$). On the other hand, pleiotropy also features in the estimating equations as error terms, and hence they contribute to the asymptotic variance of our estimators. We make the following assumption on $\delta_j$ and $\alpha_j$, which controls their asymptotic order. This assumption helps characterize the large-sample limit of the inverse matrix $\hat{\mathsf{M}}^{-1}$ as well as establish the asymptotic normality of our estimators. 
	
\begin{assumption}  [\normalfont Balanced pleiotropy] \label{Assumption-3: Balanced pleiotropy}\ \\
The pleiotropic effects, $\alpha_1,\dots, \alpha_p$, and $\delta_1, \dots, \delta_p$, are (i) mutually independent, (ii) have a zero mean: $\Expectation[\alpha_j] = \Expectation[\delta_j] = 0$ and (iii) have bounded fourth moments: $\Expectation[|\alpha_j|^4], \Expectation[|\delta_j|^4]\leq C\nu^4 \kappa^2$, where $C$ is some constant that does not depend on $j$ or $p$. 
\end{assumption}

Building on Assumptions \ref{Assumption-1: Estimated association effects}--\ref{Assumption-3: Balanced pleiotropy}, we present two useful lemmas, which highlight a few key aspects of our proposed method that differ from existing work in the literature as well as the theoretical considerations behind these differences. Proofs, additional auxiliary results, and further discussions are collected in the Supplementary Material. 

In the lemma below, we show that the bias-corrected estimators are immune to both winner's and loser's curse biases, meaning that $\hat{\beta}_{X_j,\mathtt{bc}}$ and  $\hat{\beta}_{M_j,\mathtt{bc}}$ are unbiased conditioning on the full selection events $\mathcal{S}_{\mathtt{x}}$ and $\mathcal{S}_{\mathtt{m}}$. In other words, they are immune to selection bias regardless of whether the SNP $j$ is selected. We note that this double unbiasedness is crucial for mediation analysis, as a SNP can be selected for instrumenting the exposure but not the mediator, and vice versa. 

\begin{lem}\label{lem: RB unbiasedness and variance validity} 
Let Assumptions \ref{Assumption-1: Estimated association effects} and \ref{Assumption-2:IV selection and strength } hold. Then
\begin{align*}
\Expectation[\hat{\beta}_{X_j,\mathtt{bc}}|\mathcal{S}_{\mathtt{x}},\mathcal{S}_{\mathtt{m}}] &= \beta_{X_j},\quad\text{and }
\Expectation[\hat{\beta}_{M_j,\mathtt{bc}}|\mathcal{S}_{\mathtt{x}},\mathcal{S}_{\mathtt{m}}] = \beta_{M_j}.
\end{align*}   
In addition, let ${\varsigma}_{X_j} = \Var[\hat{\beta}_{X_j,\mathtt{bc}}| \mathcal{S}_{\mathtt{x}}, \mathcal{S}_{\mathtt{m}}]$ and ${\varsigma}_{M_j}= \Var[\hat{\beta}_{M_j,\mathtt{bc}}|\mathcal{S}_{\mathtt{x}},  \mathcal{S}_{\mathtt{m}}]$. Then
\begin{align*}
   & \Big|\sum\limits_{j\in\mathcal{S}_{\mathtt{x}}}(\hat{\varsigma}_{X_j}-{\varsigma}_{X_j})\Big| \precsimProb \nu^2\sqrt{\mathsf{p}},\quad \text{and}\quad \Big|\sum\limits_{j\in\mathcal{S}_{\mathtt{m}}}(\hat{\varsigma}_{M_j}-{\varsigma}_{M_j})\Big|\precsimProb\nu^2\sqrt{\mathsf{p}}.
\end{align*} 
\end{lem}

Lemma \ref{lem: RB unbiasedness and variance validity} also provides an error bound on the estimated conditional variance of our instrument effect estimators $\hat{\beta}_{X_j,\mathtt{bc}}$ and  $\hat{\beta}_{M_j,\mathtt{bc}}$. This result is crucial for establishing the validity of our measurement removal step. 

As mentioned earlier in Sections \ref{sec2:set-up} and \ref{Sec3:Proposed-framework}, MAGIC employs carefully crafted estimating equations Eq \eqref{eq:indirect effect est equation}--Eq \eqref{eq:direct effect est equation2} to address the imperfect IV selection issue. These estimating equations not only avoid the potential bias issue induced by missing relevant IVs, but also improve the estimation efficiency by excluding irrelevant ones for estimating different causal parameters. This is in stark contrast to multivariable MR which, in essence, performs a multiple linear regression using the union of relevant IVs of the exposure and mediator. In fact, as we show in Theorem \ref{thm:asy efficiency gain} and in our simulation results Table \ref{tab: Variance for oracle scenario}, MAGIC leads to efficiency gains over multivariable MR in important settings. 

However, these carefully constructed estimating equations also make large-sample properties of the inverse matrix $\hat{\mathsf{M}}^{-1}$ considerably more difficult to establish. To gain some insight into this theoretical challenge, we first establish in the Supplementary material that
\begin{align*}
\Vert \hat{\mathsf{M}} - \mathsf{M}\Vert &\precsimProb \sqrt{\mathsf{p}\kappa},\quad \text{where }\mathsf{M}= \left[\begin{array}{ccc}
 \sum\limits_{j\in\mathcal{S}_{\mathtt{x}}}\frac{\beta_{X_j}^2}{\sigma_{Y_j}^2 }  & \sum\limits_{j\in\mathcal{S}_{\mathtt{x}}}\frac{\beta_{X_j}\beta_{M_j}}{\sigma_{Y_j}^2}&0 \\[12pt]
 \sum\limits_{j\in\mathcal{S}_{\mathtt{m}}}\frac{\beta_{M_j}\beta_{X_j}}{\sigma_{Y_j}^2 }  & \sum\limits_{j\in\mathcal{S}_{\mathtt{m}}}\frac{\beta_{M_j}^2}{\sigma_{Y_j}^2}&0 \\[12pt]
0&0&\sum\limits_{j\in\mathcal{S}_{\mathtt{x}}}\frac{\beta_{X_j}^2}{\sigma_{M_j}^2}
\end{array}\right].
\end{align*}
The next step is to provide a probabilistic order for the ``limiting matrix'' $\mathsf{M}$. To be even more precise, our goal is to show that $\mathsf{M}$ is of order $\mathsf{p}\kappa$, which is equivalent to saying that $\frac{1}{\mathsf{p}\kappa}\mathsf{M}$ is asymptotically bounded and invertible. This turns out to be a nontrivial task, because the decomposition $\beta_{M_j} = \tau_X\beta_{X_j} + \delta_j$ suggests that entries in the second row and the second column of ${\mathsf{M}}$ will depend on quantities such as $\beta_{X_j}^2$, $\beta_{X_j}\delta_j$, $\delta_j^2$, and two different selection events $\mathcal{S}_{\mathtt{x}}$ and $\mathcal{S}_{\mathtt{m}}$.  Furthermore, while $\beta_{X_j}$ and $\delta_j$ are ``uncorrelated'' under $\mathcal{S}_{\mathtt{x}}$ (c.f. our balanced horizontal pleiotropy condition in Assumption \ref{Assumption-3: Balanced pleiotropy}), the post-selection distribution of $\delta_j$ under the selection event $j \in \mathcal{S}_{\mathtt{m}}$ is no longer unrelated to $\beta_{X_j}$. In short, the asymmetry in $\hat{\mathsf{M}}$ and $\mathsf{M}$, as the result of using different instruments for the exposure and the mediator, makes it quite challenging to find the exact magnitude of the entries in those matrices, and hence their invertibility. With a careful analysis, we are able to establish an asymptotic order for $\hat{\mathsf{M}}$ and its asymptotic invertibility under proper scaling. As a by-product, we provide an auxiliary lemma in the Supplementary Material (Lemma 11), which can be employed to uncover the sign of post-selection correlations and may be of independent interest.

\begin{lem}\label{lem: inversion of M-hat}
Let Assumptions \ref{Assumption-1: Estimated association effects}--\ref{Assumption-3: Balanced pleiotropy} hold. Then $\frac{1}{\mathsf{p}\kappa}\mathsf{M}$ is bounded, and its minimum singular value is bounded away from 0. In addition,
\begin{align*}
\mathsf{M}^{-1}\hat{\mathsf{M}} \toProb \mathsf{I},
\end{align*}
where $\mathsf{I}$ is the identity matrix.  
\end{lem}

Collectively, Lemmas \ref{lem: RB unbiasedness and variance validity} and \ref{lem: inversion of M-hat} justify the following large-sample representation of our estimators: 
\begin{align*}
    \left[\begin{array}{c}
		\widehat{\theta} -\theta \\
		\widehat{\tau}_{Y}-\tau_Y\\
  \widehat{\tau}_{X}-\tau_X
	\end{array}\right] &= \hat{\mathsf{M}}^{-1} \sum\limits_{j=1}^pU_j \approx {\mathsf{M}}^{-1} \sum\limits_{j=1}^pU_j,
\end{align*}
where $\approx$ means ``equivalent up to negligible terms.'' To conserve space, we refer interested readers to the Supplementary Material for expressions of $U_j$. 

The next two assumptions are needed for the asymptotic normality result. To be precise, Assumption \ref{Assumption-4: No dominating instrument} requires that there is no dominating instrument. It is quite mild, as this assumption only rules out the (unlikely) scenario in which a few instruments are very strong while all the others are weak/irrelevant. Assumption \ref{Assumption-5: No perfect correlation} is a regularity condition, which excludes the scenario that the estimated effects are perfectly correlated (i.e., some linear combinations thereof have a zero asymptotic variance). 

\begin{assumption}[\normalfont No dominating instrument]\label{Assumption-4: No dominating instrument}
The instrument strengths satisfy:
\begin{align*}
\frac{\max\limits_{j \in \mathcal{S}_{\mathtt{x}}\cup \mathcal{S}_{\mathtt{m}}} \beta_{X_j}^2 }{\sum\limits_{j \in \mathcal{S}_{\mathtt{x}}} \beta_{X_j}^2}\toProb 0,\quad \text{and}\quad 
\frac{\max\limits_{j \in \mathcal{S}_{\mathtt{x}}\cup\mathcal{S}_{\mathtt{m}}} \beta_{M_j}^2 }{\sum\limits_{j \in \mathcal{S}_{\mathtt{m}}} \beta_{M_j}^2}\toProb 0.
\end{align*}
\end{assumption} 

\begin{assumption}[\normalfont No perfect correlation]\label{Assumption-5: No perfect correlation}
Define $\mathsf{U} = \sum_{j=1}^p \Expectation[U_jU_j^\intercal\ |\ \mathcal{S}_{\mathtt{x}},\mathcal{S}_{\mathtt{m}}]$. The minimum singular value of $\frac{1}{\mathsf{p}\kappa^2}{\mathsf{U}}$ is bounded away from 0. 
\end{assumption}

\subsection{Asymptotic normality and consistent variance estimator}

The theorem below, which is the main result of the manuscript, establishes the asymptotic distribution of our estimators, and also shows the validity of the estimated variance-covariance matrix. 

\begin{thm}\label{thm: variance consistency}
Let Assumptions \ref{Assumption-1: Estimated association effects}--\ref{Assumption-5: No perfect correlation} hold. Then conditional on the selection events,
\begin{align*}
\hat{\mathsf{V}}^{-\frac{1}{2}}\left[\begin{array}{c}
		\widehat{\theta} -\theta \\
		\widehat{\tau}_{Y}-\tau_Y\\
  \widehat{\tau}_{X}-\tau_X
	\end{array}\right] \toDist \mathcal{N}(0,\mathsf{I}),
\end{align*}
where $\mathsf{I}$ is the identity matrix. 
\end{thm}

The joint asymptotic normality characterization of the three estimated effects, $\hat{\theta}$, $\hat{\tau}_Y$, and $\hat{\tau}_X$, is another key feature of the proposed method. Together with the valid variance (covariance) estimator, they allow conducting statistical inference under both linear and nonlinear transformations of the estimators, such as the estimated mediation effect $\hat{\tau}_X\hat{\tau}_Y$ and the total effect $\hat{\theta} + \hat{\tau}_X\hat{\tau}_Y$. In addition, the estimated variance is constructed using the empirical analogue $\hat{U}_i$, whose specific form is motivated by a ``regression error'' representation of $U_i$. In other words, our variance estimator automatically incorporates the randomness in the estimated instrument associations ($\hat{\beta}_{X_j,\mathtt{bc}}$ and $\hat{\beta}_{M_j,\mathtt{bc}}$), the estimated bias correction for $\hat{\beta}_{X_j,\mathtt{bc}}^2$ and $\hat{\beta}_{M_j,\mathtt{bc}}^2$ ($\hat{\varsigma}_{X_j}$ and $\hat{\varsigma}_{M_j}$), and the pleiotropic effects ($\delta_j$ and $\alpha_j$), avoiding the need to estimate other intermediate nuisance quantities. 

As a corollary, the estimated mediation effect also has an asymptotically normal distribution, and its variance estimate is valid. 

\begin{coro}
Let Assumptions \ref{Assumption-1: Estimated association effects}--\ref{Assumption-5: No perfect correlation} hold. Then
\begin{align*}
\frac{\hat\tau-\tau_X\tau_Y}{\sqrt{\hat{\mathsf{v}}_{\tau}}} \toDist\mathcal{N}(0,1).
\end{align*}
\end{coro}

\subsection{Asymptotic efficiency gain compared to MVMR }\label{Sec:asymptotic-efficiency-gain}

In Introduction and Section \ref{Section: Method}, we initially provided a heuristic argument for the efficiency gain of the proposed method relative to MVMR, as our estimating equations allow the use of different sets of instruments for the exposure and the mediator. We now present a theoretical justification of such efficiency claims within an ``oracle" framework. The oracle setting refers to the case where we have perfect knowledge of the relevant IVs of the mediator and exposure variables in the population. That is, both $\mathcal{S}_{\mathtt{x}}^*=\left\{j:\beta_{X_j}\neq 0\right\}$ and $\mathcal{S}_{\mathtt{m}}^*=\left\{j:\beta_{M_j}\neq 0\right\}$ are known.
In this oracle framework, we show that MAGIC can provide more efficient estimators of causal effects when compared to the measurement error bias-corrected multivariable Mendelian randomization approach (DMVMR). Here, we compare MAGIC with DMVMR, rather than MVMR, because the standard MVMR still suffers from the measurement error bias. 

In this oracle setting, MAGIC and DMVMR can be written as
\begin{align*}
\left[\begin{array}{c}
		\widehat{\theta}^* \\
		\widehat{\tau}_{Y}^*
	\end{array}\right]=\left[\begin{array}{cc}
		\sum\limits_{j\in \mathcal{S}_{\mathtt{x}}^*}\frac{\widehat{\beta}_{X_j}^2-\sigma_{X_j}^2}{\sigma_{Y_j}^2} & \sum\limits_{j\in \mathcal{S}_{\mathtt{x}}^*} \frac{\widehat{\beta}_{X_j} \widehat{\beta}_{M_j}}{\sigma_{Y_j}^2} \\[15pt]
		\sum\limits_{j\in \mathcal{S}_{\mathtt{m}}^*}\frac{ \widehat{\beta}_{X_j} \widehat{\beta}_{M_j}}{\sigma_{Y_j}^2} & \sum\limits_{j\in \mathcal{S}_{\mathtt{m}}^*}\frac{\widehat{\beta}_{M_j}^2-\sigma_{M_j}^2}{\sigma_{Y_j}^2}
	\end{array}\right]^{-1}\left[\begin{array}{c}
		\sum\limits_{j\in \mathcal{S}_{\mathtt{x}}^*} \frac{\widehat{\beta}_{Y_j} \widehat{\beta}_{X_j}}{\sigma_{Y_j}^2} \\[15pt]
		\sum\limits_{j\in \mathcal{S}_{\mathtt{m}}^*}\frac{\widehat{\beta}_{Y_j} \widehat{\beta}_{M_j}}{\sigma_{Y_j}^2}
	\end{array}\right],
\end{align*}
and
\begin{align*}
\left[\begin{array}{c}
		\widehat{\theta}_{\mathtt{DMVMR}}^* \\
		\widehat{\tau}_{Y,\mathtt{DMVMR}}^* 
	\end{array}\right]=\left[\begin{array}{cc}
		\sum\limits_{j\in \mathcal{S}_{\mathtt{x}^*}\cup \mathcal{S}_{\mathtt{m}}^*}\frac{\widehat{\beta}_{X_j}^2-\sigma_{X_j}^2}{\sigma_{Y_j}^2} & \sum\limits_{j\in \mathcal{S}_{\mathtt{x}}^*\cup \mathcal{S}_{\mathtt{m}}^*} \frac{\widehat{\beta}_{X_j} \widehat{\beta}_{M_j}}{\sigma_{Y_j}^2} \\[15pt]
		\sum\limits_{j\in \mathcal{S}_{\mathtt{x}}^*\cup \mathcal{S}_{\mathtt{m}}^*} \frac{\widehat{\beta}_{X_j} \widehat{\beta}_{M_j}}{\sigma_{Y_j}^2} & \sum\limits_{j\in \mathcal{S}_{\mathtt{x}}^*\cup \mathcal{S}_{\mathtt{m}}^*}\frac{\widehat{\beta}_{M_j}^2-\sigma_{M_j}^2}{\sigma_{Y_j}^2}
	\end{array}\right]^{-1}\left[\begin{array}{c}
		\sum\limits_{j\in \mathcal{S}_{\mathtt{x}}^*\cup \mathcal{S}_{\mathtt{m}}^*} \frac{\widehat{\beta}_{Y_j} \widehat{\beta}_{X_j}}{\sigma_{Y_j}^2} \\[15pt]
		\sum\limits_{j\in \mathcal{S}_{\mathtt{x}}^*\cup \mathcal{S}_{\mathtt{m}}^*}\frac{\widehat{\beta}_{Y_j} \widehat{\beta}_{M_j}}{\sigma_{Y_j}^2}
	\end{array}\right].
\end{align*}
We do not discuss the estimation efficiency of ${\tau}_X$ in this section, as an estimator of $\tau_X$ is obtained through a separate structural equation. 

To conduct efficiency comparison, it is natural to consider three scenarios. First, when $\mathcal{S}_{\mathtt{x}}^* = \mathcal{S}_{\mathtt{m}}^*$, 
we can see that the two approaches deliver the same estimates, and therefore, our proposed MAGIC and the DMVMR have the same asymptotic variance (see Theorem \ref{thm:asy efficiency gain} below for a precise statement).  
The second scenario considers cases with partially overlapped instruments: \(\mathcal{S}_{\mathtt{x}}^* \backslash \mathcal{S}_{\mathtt{m}}^* \neq \emptyset\) and \(\mathcal{S}_{\mathtt{m}}^* \backslash \mathcal{S}_{\mathtt{x}}^* \neq \emptyset\). The third scenario explores the nested instrument case where the set \(\mathcal{S}_{\mathtt{m}}^*\) is a superset of \(\mathcal{S}_{\mathtt{x}}^*\), that is \(\mathcal{S}_{\mathtt{m}}^*\supsetneq \mathcal{S}_{\mathtt{x}}^*\). To simplify the theoretical comparison in the last two scenarios, we assume that the estimated instrument associations have the same variance: for each $j$,  $\sigma_{X_j}^2=\sigma_{Y_j}^2 = \sigma_{M_j}^2$. We believe this condition can be relaxed with the cost of having much lengthier derivations. 

In the second scenario, where there are overlapping relevant instruments and the sets \(\mathcal{S}_{\mathtt{x}}^*\) and \(\mathcal{S}_{\mathtt{m}}^*\) are not subsets of each other (i.e., \(\mathcal{S}_{\mathtt{x}}^* \backslash \mathcal{S}_{\mathtt{m}}^* \neq \emptyset\) and \(\mathcal{S}_{\mathtt{m}}^* \backslash \mathcal{S}_{\mathtt{x}}^*\neq \emptyset\)), we formally assume that:

\begin{assumption}[\normalfont Balanced instruments]\label{assu: balanced instruments}
The cardinalities of $\mathcal{S}_{\mathtt{m}}^* \backslash \mathcal{S}_{\mathtt{x}}^*$ and $\mathcal{S}_{\mathtt{x}}^* \backslash \mathcal{S}_{\mathtt{m}}^*$ satisfy
\begin{align*}
\frac{1}{6.85}\leq \frac{|\mathcal{S}_{\mathtt{m}}^* \backslash \mathcal{S}_{\mathtt{x}}^*|}{|\mathcal{S}_{\mathtt{x}}^* \backslash \mathcal{S}_{\mathtt{m}}^*|}\leq 6.85. 
\end{align*}
\end{assumption}
\noindent The threshold, $6.85$, is solved from a polynomial equation. (See Theorem 2 in the Supplementary Material and its proof for details.) 

In the third scenario, where the instruments are nested with $\mathcal{S}_{\mathtt{m}}^*\supsetneq \mathcal{S}_{\mathtt{x}}^*$, we impose the following assumption. Here, with slight abuse of notation, we use $p_{\mathtt{x}}, p_{\mathtt{m}}, \kappa_{\mathtt{x}}, \kappa_{\mathtt{m}}^{\delta}$ to denote the counterpart of notations described in Section \ref{Sec:notations} when the selected SNP sets are oracle $\mathcal{S}_{\mathtt{x}}^*$ and  $\mathcal{S}_{\mathtt{m}}^*$.

\begin{assumption}[\normalfont Instrument strengths differential] \label{Assump-7 instrument strengths differential} For SNPs in  $\mathcal{S}_{\mathtt{x}}^*$ and $\mathcal{S}_{\mathtt{m}}^*$ and the effects $(\theta, \tau_Y, \tau_X)$,
\begin{enumerate}
    \item[(i)]The instrument strengths  satisfy   $8p_{\mathtt{m}}\kappa^{\delta}_{\mathtt{m}} > p_{\mathtt{x}}\kappa_{\mathtt{x}}$.
    \item[(ii)] The effects satisfy $\tau_X^2  \Big(1+ 2\theta^2+\tau_Y^2\Big) > 2 \tau_X \theta\tau_Y$. 
\end{enumerate}
\end{assumption}
\noindent We note that this scenario can be quite relevant for mediation analysis, as whenever $\tau_X \neq 0$, a SNP relevant for $X$ is likely relevant to $M$ according to the structural equation ${\beta}_{M_j} =\beta_{X_j} \cdot \tau_X \cdot\Indicator_{(j\in  \mathcal{S}_{\mathtt{x}}^*)} +\delta_j$. We have provided a more detailed demonstration of Assumption \ref{Assump-7 instrument strengths differential}(ii) in the Supplementary Material Section 6.

We present in the theorem below the asymptotic efficiency comparison results. We will use $\mathsf{V}^*$ to denote the asymptotic variance-covariance matrix of $\widehat{\theta}^*$ and $\widehat{\tau}_{Y}^*$, and use $\mathsf{V}_{\mathtt{DMVMR}}^*$ for DMVMR. Diagonal elements of $\mathsf{V}^*$ are denoted as $\mathsf{v}_{\theta}^*$ and $\mathsf{v}_{\tau_Y}^*$. Similarly, $\mathsf{v}_{\theta,\mathtt{DMVMR}}^*$ and $\mathsf{v}_{\tau_Y,\mathtt{DMVMR}}^*$ are the diagonal entries of $\mathsf{V}_{\mathtt{DMVMR}}^*$. 
\begin{thm} \label{thm:asy efficiency gain} In the oracle setting, 
\begin{enumerate}
    \item[(1)] Complete overlapping with $\mathcal{S}_{\mathtt{x}}^* = \mathcal{S}_{\mathtt{m}}^*$: MAGIC and DMVMR have the same asymptotic variance matrix, $\mathsf{V}^*=\mathsf{V}_{\mathtt{DMVMR}}^*$.
    \item[(2)] Partial overlapping with $\mathcal{S}_{\mathtt{x}}^* \backslash \mathcal{S}_{\mathtt{m}}^* \neq \emptyset$ and $\mathcal{S}_{\mathtt{m}}^* \backslash \mathcal{S}_{\mathtt{x}}^* \neq \emptyset$: under Assumption \ref{assu: balanced instruments},  $\mathsf{V}_{\mathtt{DMVMR}}^* - \mathsf{V}^*$ is positive definite.
    \item[(3)] Nested IVs with $\mathcal{S}_{\mathtt{m}}^*\supsetneq \mathcal{S}_{\mathtt{x}}^*$: Under  Assumption \ref{Assump-7 instrument strengths differential}(i), $\mathsf{v}_{\theta, \mathtt{DMVMR}}^* > \mathsf{v}_{\theta}^*$; under Assumption \ref{Assump-7 instrument strengths differential}(ii), $\mathsf{v}_{\tau_Y, \mathtt{DMVMR}}^* > \mathsf{v}_{\tau_Y}^*$.
\end{enumerate}
\end{thm}

The theorem provides sufficient conditions under which MAGIC is asymptotically more efficient than DMVMR. Although we are not able to show that MAGIC is universally more efficient under all possible data generating processes, we note that the assumptions (\ref{assu: balanced instruments} and \ref{Assump-7 instrument strengths differential}) are quite mild and are likely to hold in realistic mediation analysis settings. In addition, the conditions we impose in the theorem are sufficient, but by no means necessary: this is also borne out in our simulation studies, as our approach demonstrates efficiency gain over DMVMR even if Assumptions \ref{assu: balanced instruments} and \ref{Assump-7 instrument strengths differential} do not hold. 

\section{Simulation studies}
\label{Sec:simulation-study}

In this section, we compare the performance of MAGIC with several other methods: a direct plug-in estimator, multivariable MR, debiased multivariable MR, and two-step MR. 

\subsection{Simulation design}\label{Sec:simulation-design}

Following the notation adopted in the previous section, let $\mathcal{S}_{\mathtt{x}}^* = \{j: \beta_{X_j}\neq 0\}$, $\mathcal{S}_{\mathtt{m}}^* = \{j: \beta_{M_j}\neq 0\}$, and $\mathcal{S}_{\delta}^*=\{j: \delta_{j}\neq 0\}$. We consider three data-generating processes (DGPs), and their main difference is the fraction of generic instruments shared by the exposure and the mediation: \textbf{DGP 1}. Complete overlapping with $\mathcal{S}_{\mathtt{x}}^* = \mathcal{S}_{\mathtt{m}}^*$ generated by $\tau_X = 0.6$, and $\mathcal{S}_{\mathtt{x}}^* =\mathcal{S}_{\delta}^*$. \textbf{DGP 2}. Partial overlapping with $\mathcal{S}_{\mathtt{x}}^* \backslash \mathcal{S}_{\mathtt{m}}^* \neq \emptyset$ and $\mathcal{S}_{\mathtt{m}}^* \backslash \mathcal{S}_{\mathtt{x}}^* \neq \emptyset$: $\tau_X = 0$ with either (i) $\mathcal{S}_{\mathtt{x}}^*  \cap \mathcal{S}_{\delta}^* = \emptyset$,  or (ii) $| \mathcal{S}_{\mathtt{x}}^*  \cap \mathcal{S}_{\delta}^*| = | \mathcal{S}_{\mathtt{x}}^*| /2 $. \textbf{DGP 3}. Nested IVs with $\mathcal{S}_{\mathtt{m}}^*\supsetneq \mathcal{S}_{\mathtt{x}}^*$:  $\tau_X = 0.6$ with either (i) $\mathcal{S}_{\mathtt{x}}^*  \cap \mathcal{S}_{\delta}^* = \emptyset$,  or (ii) $| \mathcal{S}_{\mathtt{x}}^*  \cap \mathcal{S}_{\delta}^*| = | \mathcal{S}_{\mathtt{x}}^*| /2 $. See Section 7.1 in  the Supplementary Material for additional details. 

Having specified the relevant instruments, we next generate the true GWAS association effects as: 
\begin{align*}
    \text{for $j \in \mathcal{S}_{\mathtt{x}}^*$},\quad \beta_{X_j} \sim \mathcal{N}(0, \varepsilon_{\mathtt{x}}^2);\qquad \text{for $j \in \mathcal{S}_{\delta}^*$},\quad \delta_j  \sim \mathcal{N}(0, \varepsilon_{\delta}^2).
\end{align*}
To control the proportion of relevant IVs, we
use $\pi_{\mathtt{x}}$ to denote the proportion of SNPs with $\beta_{X_j}\neq 0$, and $\pi_{\delta}$ to denote the proportion of SNPs with $\delta_j\neq 0$ (that is, $\pi_{\mathtt{x}} = |\mathcal{S}_{\mathtt{x}}^*| / p$ and $\pi_{\delta} = |\mathcal{S}_{\delta}^*| / p$). As for the effect size of $\beta_{M_j}$ and $\beta_{Y_j}$, following the causal diagram in Figure \ref{fig:diagram}, we construct $\beta_{M_j}=\tau_X\beta_{X_j} + \delta_j, \text{and } \beta_{Y_j} =\theta\beta_{X_j} +\tau_Y\beta_{M_j}$. Without loss of generality, we also consider a scenario in which the standard deviations of the measured associations are equivalent across different GWAS summary statistics with $\sigma_{X_j}^2 = \sigma_{Y_j}^2 = \sigma_{M_j}^2=1/100,000$. Following the above data-generating processes, we generate 1,000 Monte Carlo samples in each simulation design. 

To showcase the merit of the proposed estimator, we compare it with four methods, including ``plug-in,'' ``MVMR,'' ``DMVMR'' and ``Two-step MR.'' Plug-in estimator solves the sample analog of Eq \eqref{eq:structural-eq-oracle1}-\eqref{eq:structural-eq-oracle2}, and can be expressed as 
{\footnotesize\begin{align*}
\begin{bmatrix}
\hat{\theta}^{\mathtt{PI}} \\
\hat{\tau}_Y^{\mathtt{PI}} \\
\hat{\tau}_X^{\mathtt{PI}}
\end{bmatrix} &= 
\begin{bmatrix}
\sum\limits_{j \in \mathcal{S}_{\mathtt{x}}}{(\widehat{\beta}_{X_j, \mathtt{bc}}^2-\hat{\varsigma}_{X_j})}/{\sigma_{Y_j}^2} & \sum\limits_{j \in \mathcal{S}_{\mathtt{x}}\cap \mathcal{S}_{\mathtt{m}}} {\widehat{\beta}_{X_j, \mathtt{bc}} \widehat{\beta}_{M_j, \mathtt{bc}} }/{\sigma_{Y_j}^2}&0 \\[15pt]
		\sum\limits_{j \in \mathcal{S}_{\mathtt{x}}\cap\mathcal{S}_{\mathtt{m}}} {\widehat{\beta}_{X_j, \mathtt{bc}} \widehat{\beta}_{M_j, \mathtt{bc}}}/{\sigma_{Y_j}^2}  & \sum\limits_{j \in \mathcal{S}_{\mathtt{m}}}{(\widehat{\beta}_{M_j, \mathtt{bc}}^2-\hat{\varsigma}_{M_j})}/{\sigma_{Y_j}^2}&0\\[15pt]
  0&0&\sum\limits_{j \in \mathcal{S}_{\mathtt{x}}}{(\widehat{\beta}_{X_j, \mathtt{bc}}^2-\hat{\varsigma}_{X_j})}/{\sigma_{M_j}^2} 
\end{bmatrix}^{-1}
\begin{bmatrix}
\sum\limits_{j \in \mathcal{S}_{\mathtt{x}}} {\widehat{\beta}_{X_j, \mathtt{bc}}\widehat{\beta}_{Y_j} }/ { \sigma_{Y_j}^2 }\\[15pt]
		\sum\limits_{j \in \mathcal{S}_{\mathtt{m}}} {\widehat{\beta}_{M_j, \mathtt{bc}} \widehat{\beta}_{Y_j} }/{\sigma_{Y_j}^2}\\[15pt]
  \sum\limits_{j \in \mathcal{S}_{\mathtt{x}}} {  \widehat{\beta}_{X_j, \mathtt{bc}}\widehat{\beta}_{M_j, \mathtt{bc}} }/{ \sigma_{M_j}^2}
\end{bmatrix}.
\end{align*}
}
Using the plug-in approach, the mediation effect can be obtained through  $\hat{\tau}^{\mathtt{PI}}=\hat{\tau}_X^{\mathtt{PI}}\hat{\tau}_Y^{\mathtt{PI}}$. We note that the plug-in estimator is immune to the winner's curse and the measurement error bias. However, it suffers from the imperfect IV selection bias.

Next, as the MVMR, DMVMR and two-step MR methods do not address the winner's curse bias issue, they select relevant IVs using a hard threshold-crossing rule:
\begin{align*}
    &\tilde{\mathcal{S}}_{\mathtt{x}}=\Big\{j:\Big| \frac{\hat{\beta}_{X_j}}{ \sigma_{X_j} } \Big| > \lambda, \ j=1, \ldots, p\Big\} \text{ and } \tilde{\mathcal{S}}_{\mathtt{m}}=\Big\{j:\Big| \frac{\hat{\beta}_{M_j}}{ \sigma_{M_j} } \Big| > \lambda, \ j=1, \ldots, p\Big\}.
\end{align*} 
MVMR can be viewed as a multivariable linear regression using associations \citep{dyy262}: 
\begin{align*}
\small \left[\begin{array}{c}
     \hat{\theta}_{\mathtt{MVMR}}\\
   \hat{\tau}_{Y, \mathtt{MVMR}}
\end{array}\right] =   \begin{bmatrix}
		\sum\limits_{j\in \tilde{\mathcal{S}}_{\mathtt{x}}\cup \tilde{\mathcal{S}}_{\mathtt{m}}  }{\widehat{\beta}_{X_j}^2}/ { \sigma_{Y_j}^2} & \sum\limits_{j\in \tilde{\mathcal{S}}_{\mathtt{x}}\cup \tilde{\mathcal{S}}_{\mathtt{m}} } {\widehat{\beta}_{X_j} \widehat{\beta}_{M_j}}/{ \sigma_{Y_j}^2} \\
		\sum\limits_{j \in \tilde{\mathcal{S}}_{\mathtt{x}}\cup \tilde{\mathcal{S}}_{\mathtt{m}} }{\widehat{\beta}_{X_j} \widehat{\beta}_{M_j}}/{ \sigma_{Y_j}^2} & \sum\limits_{j\in \tilde{\mathcal{S}}_{\mathtt{x}}\cup \tilde{\mathcal{S}}_{\mathtt{m}} }{\widehat{\beta}_{M_j}^2}/{ \sigma_{Y_j}^2}
\end{bmatrix}^{-1}\begin{bmatrix}
		\sum\limits_{j\in\tilde{\mathcal{S}}_{\mathtt{x}}\cup \tilde{\mathcal{S}}_{\mathtt{m}} } {\widehat{\beta}_{Y_j} \widehat{\beta}_{X_j}}/{ \sigma_{Y_j}^2} \\
		\sum\limits_{j \in \tilde{\mathcal{S}}_{\mathtt{x}}\cup \tilde{\mathcal{S}}_{\mathtt{m}} }{\widehat{\beta}_{Y_j} \widehat{\beta}_{M_j}}/{ \sigma_{Y_j}^2}
	\end{bmatrix}.
\end{align*}
DMVMR corrects the measurement error bias in  MVMR:
 \begin{align}\label{eq:mvmrdebiased}
\small \left[\begin{array}{c}
     \hat{\theta}_{\mathtt{DMVMR}}\\
   \hat{\tau}_{Y, \mathtt{DMVMR}}
\end{array}\right] = \begin{bmatrix}
		\sum\limits_{j\in \tilde{\mathcal{S}}_{\mathtt{x}}\cup \tilde{\mathcal{S}}_{\mathtt{m}}  }{(\widehat{\beta}_{X_j}^2-\sigma_{X_j}^2)}/ { \sigma_{Y_j}^2} & \sum\limits_{j\in \tilde{\mathcal{S}}_{\mathtt{x}}\cup \tilde{\mathcal{S}}_{\mathtt{m}}} {\widehat{\beta}_{X_j} \widehat{\beta}_{M_j}}/{ \sigma_{Y_j}^2} \\
		\sum\limits_{j \in \tilde{\mathcal{S}}_{\mathtt{x}}\cup \tilde{\mathcal{S}}_{\mathtt{m}}}{ \widehat{\beta}_{X_j} \widehat{\beta}_{M_j} }/{ \sigma_{Y_j}^2} & \sum\limits_{j\in \tilde{\mathcal{S}}_{\mathtt{x}}\cup \tilde{\mathcal{S}}_{\mathtt{m}}}{(\widehat{\beta}_{M_j}^2-\sigma_{M_j}^2)}/{ \sigma_{Y_j}^2}
\end{bmatrix}^{-1}\begin{bmatrix}
		\sum\limits_{j\in \tilde{\mathcal{S}}_{\mathtt{x}}\cup \tilde{\mathcal{S}}_{\mathtt{m}}} {\widehat{\beta}_{Y_j} \widehat{\beta}_{X_j}}/{ \sigma_{Y_j}^2} \\
		\sum\limits_{j\in\tilde{\mathcal{S}}_{\mathtt{x}}\cup \tilde{\mathcal{S}}_{\mathtt{m}}}{\widehat{\beta}_{Y_j} \widehat{\beta}_{M_j}}/{ \sigma_{Y_j}^2}
	\end{bmatrix}.
\end{align}

Lastly, two-step MR estimates the causal effects with a two-step procedure:
\begin{align*}
\small
    \hat{\tau}_{X, \mathtt{2-step}}=\frac{\sum\limits_{j\in \tilde{\mathcal{S}}_{\mathtt{x}}}\hat{\beta}_{M_j}\hat{\beta}_{X_j}/\sigma_{M_j}^2}{\sum\limits_{j\in \tilde{\mathcal{S}}_{\mathtt{x}}}\hat{\beta}_{X_j}^2/\sigma_{M_j}^2}, \quad \hat{\tau}_{Y, \mathtt{2-step}}=\frac{\sum\limits_{j\in \tilde{\mathcal{S}}_{\mathtt{m}}\setminus\tilde{\mathcal{S}}_{\mathtt{x}}}\hat{\beta}_{Y_j}\hat{\beta}_{M_j}/\sigma_{Y_j}^2}{\sum\limits_{j\in  \tilde{\mathcal{S}}_{\mathtt{m}}\setminus\tilde{\mathcal{S}}_{\mathtt{x}}}\hat{\beta}_{M_j}^2/\sigma_{Y_j}^2}.
\end{align*}
Similar to the MVMR approach, the two-step estimates suffer from both the winner's curse and the measurement error bias. 

In what follows, we briefly summarize implementation details of the three methods (MVMR-IVW, DMVMR, and two-step) in Table \ref{tab:methods summarization}. 
\begin{table}[H]
  \centering
    \begin{tabular}{cccccccc}
\hline
\hline
          & \multicolumn{2}{c}{MVMR-IVW} & \multicolumn{2}{c}{DMVMR} & \multicolumn{2}{c}{Two-step MR} \\
          \hline
          & Point Est. & Std. Err. & Point Est. & Std. Err. & Point Est. & Std. Err.\\
          \hline
    $\hat{\theta}$ &  $\mathtt{MVMR}$      &   $\mathtt{MVMR}$   &  Eq~\eqref{eq:mvmrdebiased}    &   N/A   &    N/A   &  N/A\\
   $\hat{\tau}_Y$ &    $\mathtt{MVMR}$  &  $\mathtt{MVMR}$     &  Eq~\eqref{eq:mvmrdebiased}    &    N/A    &  $\mathtt{TwoSampleMR}$       &$\mathtt{TwoSampleMR}$    \\
    $\hat{\tau}$   &     $\hat{\tau}_{\mathtt{xy, total}} - \hat{\theta}$   &    N/A   &   N/A    &  N/A     &   $\hat{\tau}_{Y}*\hat{\tau}_{X, \mathtt{2-step}}$    & Delta method \\
\hline
\hline
    \end{tabular}%
      \caption{Different methods implementation details. $\mathtt{R}$ package $\mathtt{MVMR}$ comes from \cite{sanderson2021testing}. $\mathtt{R}$ package $\mathtt{TwoSampleMR}$ comes from \cite{burgess2013mendelian}.  $\hat{\tau}_{\mathtt{xy, total}}$ (or $\hat{\tau}_{X, \mathtt{2-step}}$) is the estimated total effect of $X$ on $Y$ (or $X$ on $M$) obtained from the univariate MR package $\mathtt{TwoSampleMR}$. In particular, as the theoretical property of the DMVMR estimator has not been studied in the existing literature, its standard error is not available (marked as N/A in the table above). {\tablefootnote{When conducting mediation using two-step MR, $\theta$ is not the parameter of interest. Thus, we mark N/A here. To get $\hat{\tau}$'s standard error, the Delta Method can only be valid when the selected SNPs in the first step (estimating $\tau_X$) have no overlap with that of the second step (estimating $\tau_Y$). In our simulation, when it comes to the selected SNPs in the second step, we screen out the SNPs obtained at the first stage.}}}
  \label{tab:methods summarization}%
\end{table}%

 Following common practices, the cutoff value $\lambda$ is set at 5.45 (corresponding to the significance threshold $5\times 10^{-8}$) for MVMR, DMVMR, and Two-step MR estimators to select relevant IVs for both $X$ and $M$. This common practice is often conducted to avoid substantial winner's curse bias. As MAGIC and Plug-in estimator remove the winner's curse bias, we conduct them with a more liberal cutoff $\lambda=4.06$ (corresponding to the significance threshold $5\times 10^{-5}$). We set $\eta = 0.5$, and the performance of the randomized instrument based approach is not sensitive to the choice of $\eta$.

\subsection{Bias, standard deviation, power and coverage}\label{Sec:simulation-comparison} 

\begin{figure}[H]
\centering
\includegraphics[scale=0.47]{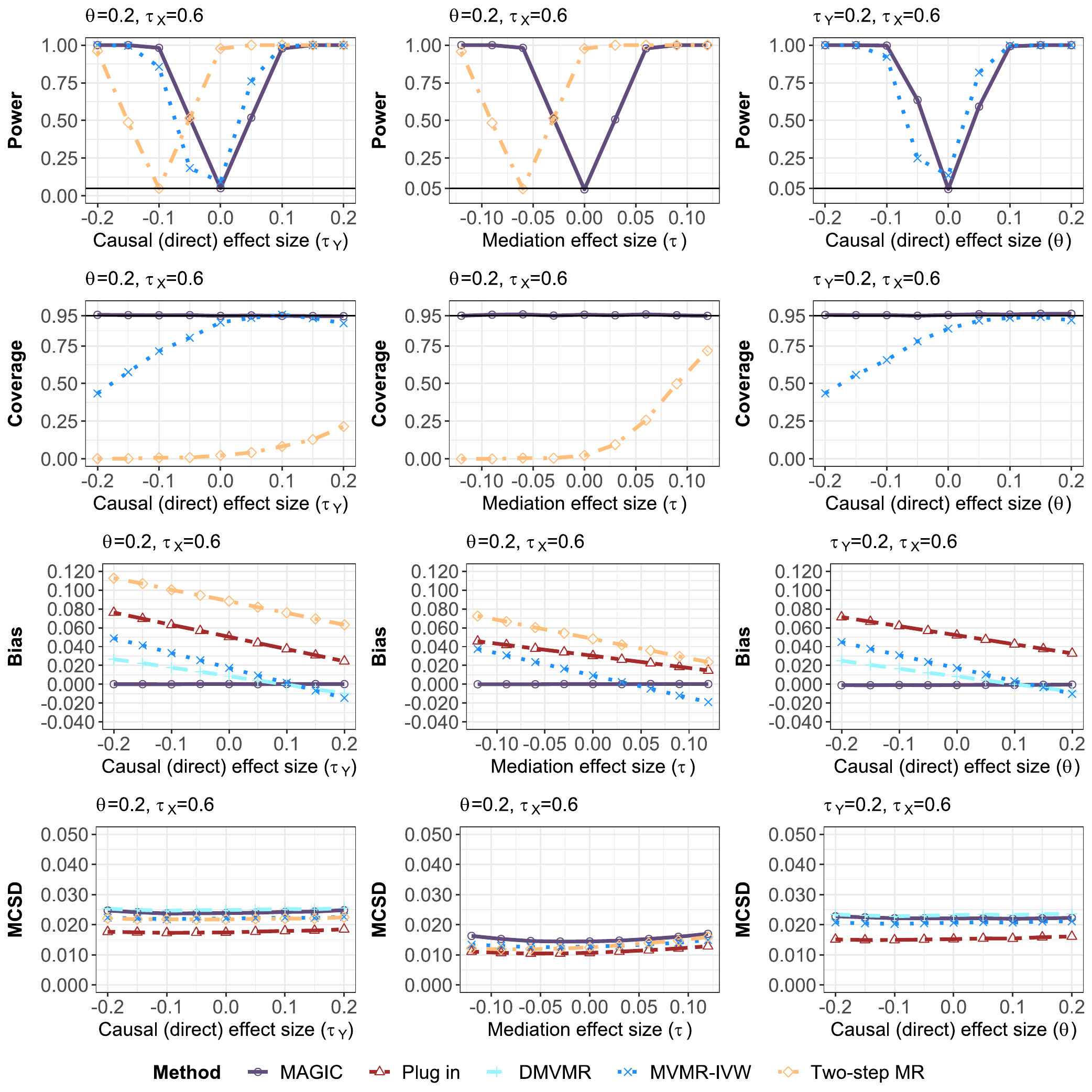}
\caption{DGP 1. Power, coverage, bias, and MCSD of MAGIC, Plug-in estimator, MVMR, DMVMR, and Two-step MR.
First and second columns are computed under $\theta=0.2, \tau_X=0.6$, $\pi_{\mathtt{x}}=\pi_{\delta}=0.01$, and $\varepsilon_{\mathtt{x}}^2=1\times 10^{-4}$, $\varepsilon_{\delta}^2=5\times 10^{-5}$, that is for the second row, $\tau \in\{-0.12, -0.09, -0.06, -0.03, 0, 0.03, 0.06, 0.09, 0.12\}$. The third column is computed when  $\tau_Y=0.2$ and $\tau_X=0.6$. As two-step MR does not estimate $\theta$, it is not contained in the third row.}\label{fig:simulation-design3-totaloverlapping}
\end{figure}

\begin{figure}[H]
\centering
\includegraphics[scale=0.47]{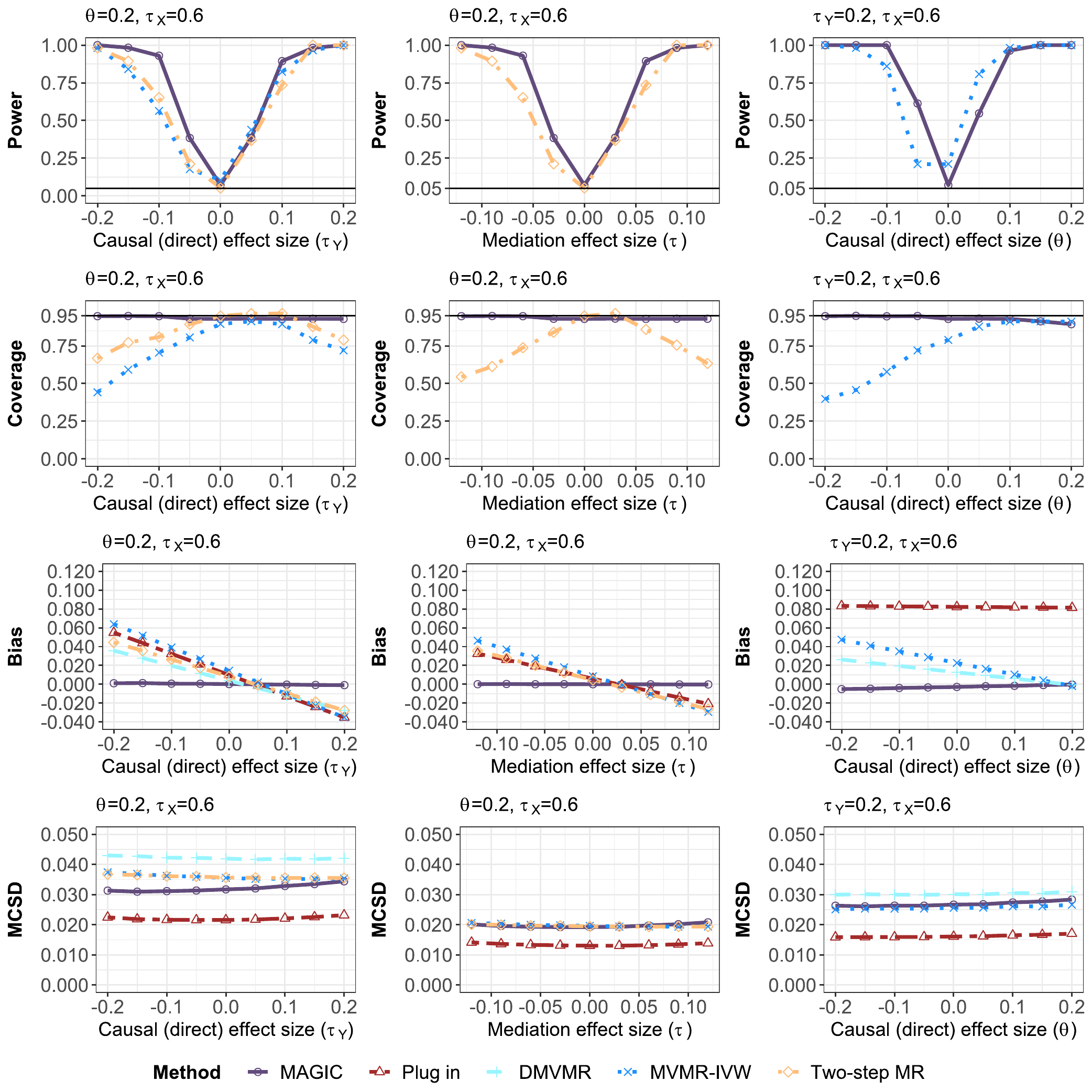}
\caption{DGP 3(i). Power, coverage, bias, and MCSD of MAGIC, Plug-in estimator, MVMR, DMVMR, and Two-step MR.
First and second columns are computed under $\theta=0.2, \tau_X=0.6$, $\pi_{\mathtt{x}}=\pi_{\delta}=0.01$, and $\varepsilon_{\mathtt{x}}^2=1\times 10^{-4}$, $\varepsilon_{\delta}^2=5\times 10^{-5}$, that is for the second row, $\tau \in\{-0.12,-0.09, -0.06, -0.03, 0, 0.03, 0.06, 0.09, 0.12\}$.The third column is computed when  $\tau_Y=0.2$ and $\tau_X=0.6$.  As two-step MR does not estimate $\theta$, it is not contained in the third row.}\label{fig:simulation-design1-nonoverlapping}
\end{figure}

\begin{figure}[H]
\centering
\includegraphics[scale=0.47]{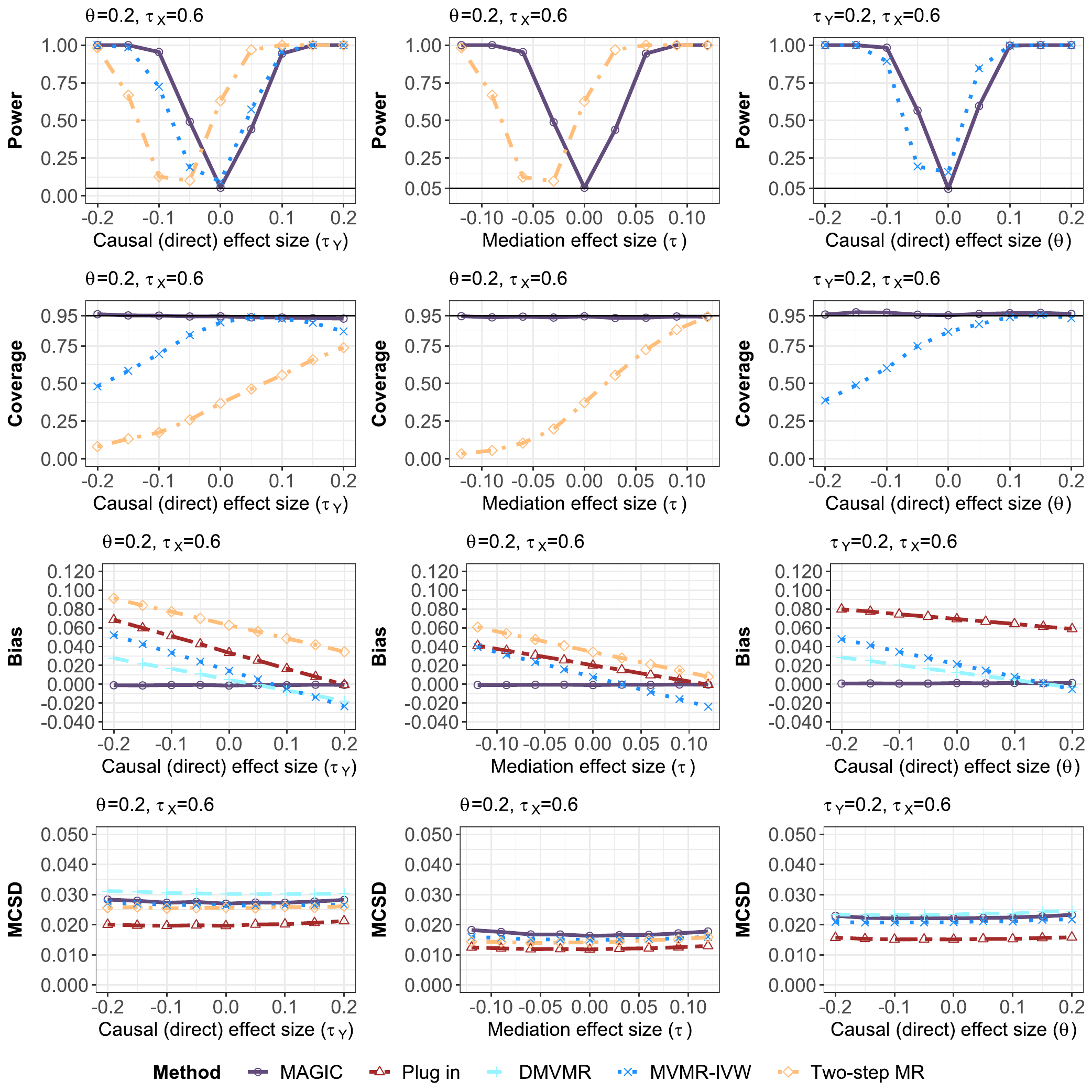}
\caption{DGP 3(ii). Power, coverage, bias, and MCSD of MAGIC, Plug-in estimator, MVMR-IVW, DMVMR, and Two-step MR.
First and second columns are computed under $\theta=0.2, \tau_X=0.6$, $\pi_{\mathtt{x}}=\pi_{\delta}=0.01$, and $\varepsilon_{\mathtt{x}}^2=1\times 10^{-4}$, $\varepsilon_{\delta}^2=5\times 10^{-5}$, that is for the second row, $\tau \in\{-0.12, -0.09, -0.06, -0.03, 0, 0.03, 0.06, 0.09, 0.12\}$. The third column is computed when  $\tau_Y=0.2$ and $\tau_X=0.6$. As two-step MR does not estimate $\theta$, it is not contained in the third row.}\label{fig:simulation-design2-halfoverlapping}
\end{figure}

Figure \ref{fig:simulation-design3-totaloverlapping} summarizes the performance of various estimators under DGP 1, and Figure \ref{fig:simulation-design1-nonoverlapping}, \ref{fig:simulation-design2-halfoverlapping} present results under DGP 3. As simulation results for DGP 2 and 3 are similar, we refer interested readers to Supplementary Section 7.3.1 for simulation results corresponding to DGP 2.  

Concretely, in each figure, we showcase the performance of the five estimators described in the previous section in terms of their ``Power'' (average rejection probability for 5\% tests), ``Coverage'' (average empirical coverage probability of 95\% confidence intervals), ``Bias'' (average difference between the estimates and the true parameters), and ``MCSD'' (Monte Carlo standard deviation). Notice that we only report Bias and MCSD if standard errors are not available.

We first focus on the plug-in approach (dotdashed burgundy) in Figure \ref{fig:simulation-design3-totaloverlapping}--\ref{fig:simulation-design2-halfoverlapping}. Across the different simulation designs, the plug-in estimators exhibit large bias due to imperfect IV selection. Again, this is because the plug-in approach employs $\mathcal{S}_{\mathtt{x}}$ and $\mathcal{S}_{\mathtt{m}}$ to empirically estimate $\mathcal{S}_{\mathtt{x}}^*$ and $\mathcal{S}_{\mathtt{m}}^*$, which are generally imperfect (and imprecise). Such imperfect IV selection biases the constructed estimating equations. 

As we discussed, MVMR-IVW (dotted blue) and two-step MR (dotdashed orange) suffer from both measurement error bias and the winner's curse. This is clear from the simulation results: estimates obtained from these two approaches are generally biased, leading to shifted power curves and poor empirical coverage. Although DMVMR (dashed pale blue) corrects the measurement error bias, it is still biased due to the winner's curse. 

Finally, we highlight that the proposed MAGIC approach (solid purple) delivers estimates that are almost unbiased without incurring efficiency loss. As a result, statistical inference using the MAGIC approach exhibits high detection power, well-controlled empirical size, and near-nominal coverage probabilities. Another interesting finding is that MAGIC typically exhibits lower Monte Carlo standard deviations compared with DMVMR, which is in line with our efficiency result in Section \ref{Sec:asymptotic-efficiency-gain}. 

To provide further evidence on the relative efficiency gain of MAGIC over DMVMR, we consider an oracle simulation setup in which the relevant SNPs, $\mathcal{S}_{\mathtt{x}}^*$ and $\mathcal{S}_{\mathtt{m}}^*$, are known a priori. In Table \ref{tab: Variance for oracle scenario}, we report the Monte Carlo standard deviations (MCSD) of MAGIC and DMVMR estimators with the ratio $|\mathcal{S}_{\mathtt{m}}^*\setminus \mathcal{S}_{\mathtt{x}}^*|/|\mathcal{S}_{\mathtt{x}}^*\setminus \mathcal{S}_{\mathtt{m}}^*|\in\{0.1, 1, 3\}$ using index construction way illustrated in Supplementary Material Section 7.2.

In line with our theoretical analysis in Theorem \ref{thm:asy efficiency gain},  we observe that the MCSD of $\widehat{\theta}^*$ and $\widehat{\tau}_Y^*$ are consistently smaller than that of  $\widehat{\theta}_{\mathtt{DMVMR}}^*$ and $\widehat{\tau}_{Y, \mathtt{DMVMR}}^*$ within the chosen ratios.  We leave results for  $|\mathcal{S}_{\mathtt{m}}^*\setminus \mathcal{S}_{\mathtt{x}}^*|/|\mathcal{S}_{\mathtt{x}}^*\setminus \mathcal{S}_{\mathtt{m}}^*|\in\{0.5, 6\}$ and $\mathcal{S}_{\mathtt{m}}^*=\mathcal{S}_{\mathtt{x}}^*$ in Supplementary Material Section 7.3.2.

\begin{table}[H]
  \centering
 
    \begin{tabular}{ccccccc}
   \hline
   \hline
          & \multicolumn{2}{c}{ ${|\mathcal{S}_{\mathtt{x}}^* \setminus\mathcal{S}_{\mathtt{m}}^*|}/{|\mathcal{S}_{\mathtt{m}}^* \setminus\mathcal{S}_{\mathtt{x}}^*|} = 0.1 $ } & \multicolumn{2}{c}{${|\mathcal{S}_{\mathtt{x}}^* \setminus\mathcal{S}_{\mathtt{m}}^*|}/{|\mathcal{S}_{\mathtt{m}}^* \setminus\mathcal{S}_{\mathtt{x}}^*|} = 1$}&\multicolumn{2}{c}{ ${|\mathcal{S}_{\mathtt{x}}^* \setminus\mathcal{S}_{\mathtt{m}}^*|}/{|\mathcal{S}_{\mathtt{m}}^* \setminus\mathcal{S}_{\mathtt{x}}^*|} = 3 $ } \\
\cmidrule{2-7}          &{DMVMR$^*$} & MAGIC$^*$ & {DMVMR$^*$ } & {MAGIC$^*$} & {DMVMR$^*$} & {MAGIC$^*$}\\
  \hline
     \multicolumn{7}{c}{$\pi_{{\mathtt{x}}\setminus {\mathtt{m}}}=\pi_{{\mathtt{m}}\setminus {\mathtt{x}}}=\pi_{{\mathtt{x}}\cap {\mathtt{m}}}=0.0005, \varepsilon_{\mathtt{x}}^2= 5\times 10^{-5}, \varepsilon_{\delta}^2=5\times 10^{-5}$} \\
    $\hat\theta$ & 0.076 & 0.055 & 0.057 & 0.054 & 0.059 & 0.058 \\
   $\hat{\tau}_Y$ & 0.022 & 0.022 & 0.052& 0.049 & 0.062 & 0.058 \\
  \hline
    \multicolumn{7}{c}{$\pi_{{\mathtt{x}}\setminus {\mathtt{m}}}=\pi_{{\mathtt{m}}\setminus {\mathtt{x}}}=\pi_{{\mathtt{x}}\cap {\mathtt{m}}}=0.0005, \varepsilon_{\mathtt{x}}^2= 5\times 10^{-5}, \varepsilon_{\delta}^2=1\times 10^{-4}$} \\
    $\hat\theta$ & 0.075 & 0.053 & 0.056 & 0.053 & 0.056 & 0.055 \\
   $\hat{\tau}_Y$ & 0.015 & 0.015  & 0.035 & 0.034 & 0.044& 0.043 \\
  \hline
    \multicolumn{7}{c}{$\pi_{{\mathtt{x}}\setminus {\mathtt{m}}}=\pi_{{\mathtt{m}}\setminus {\mathtt{x}}}=\pi_{{\mathtt{x}}\cap {\mathtt{m}}}=0.0005, \varepsilon_{\mathtt{x}}^2= 5\times 10^{-5}, \varepsilon_{\delta}^2=3\times 10^{-4}$} \\
   $\hat\theta$ & 0.073 & 0.050 & 0.057 & 0.054 & 0.055 &0.054  \\
    $\hat{\tau}_Y$ & 0.010 & 0.010 & 0.020 & 0.020 & 0.024 & 0.024 \\
     \hline
    \multicolumn{7}{c}{$\pi_{{\mathtt{x}}\setminus {\mathtt{m}}}=\pi_{{\mathtt{m}}\setminus {\mathtt{x}}}=\pi_{{\mathtt{x}}\cap {\mathtt{m}}}=0.001, \varepsilon_{\mathtt{x}}^2= 5\times 10^{-5}, \varepsilon_{\delta}^2=1\times 10^{-4}$} \\
    $\hat\theta$ & 0.050 & 0.037 & 0.039&0.036 & 0.039 & 0.038 \\
    $\hat{\tau}_Y$ & 0.010 & 0.010 & 0.025 & 0.024& 0.031 & 0.029 \\
     \hline
    \multicolumn{7}{c}{$\pi_{{\mathtt{x}}\setminus {\mathtt{m}}}=\pi_{{\mathtt{m}}\setminus {\mathtt{x}}}=\pi_{{\mathtt{x}}\cap {\mathtt{m}}}=0.001, \varepsilon_{\mathtt{x}}^2= 5\times 10^{-5}, \varepsilon_{\delta}^2=3\times 10^{-4}$} \\
    $\hat\theta$ & 0.052 & 0.037 & 0.038 & 0.036 & 0.039 & 0.038 \\
    $\hat{\tau}_Y$ & 0.006 & 0.006 & 0.013 & 0.013 & 0.016 & 0.016 \\
      \hline
    \multicolumn{7}{c}{$\pi_{{\mathtt{x}}\setminus {\mathtt{m}}}=\pi_{{\mathtt{m}}\setminus {\mathtt{x}}}=\pi_{{\mathtt{x}}\cap {\mathtt{m}}}=0.002, \varepsilon_{\mathtt{x}}^2= 5\times 10^{-5}, \varepsilon_{\delta}^2=1\times 10^{-4}$} \\
    $\hat\theta$ & 0.035 & 0.025 & 0.027 & 0.026 & 0.028 & 0.027 \\
    $\hat{\tau}_Y$ & 0.007 & 0.007 & 0.018 & 0.018 & 0.022 & 0.021 \\
      \hline
    \multicolumn{7}{c}{$\pi_{{\mathtt{x}}\setminus {\mathtt{m}}}=\pi_{{\mathtt{m}}\setminus {\mathtt{x}}}=\pi_{{\mathtt{x}}\cap {\mathtt{m}}}=0.002, \varepsilon_{\mathtt{x}}^2= 5\times 10^{-5}, \varepsilon_{\delta}^2=3\times 10^{-4}$} \\
    $\hat\theta$ & 0.035 & 0.025 & 0.027& 0.025 & 0.026 & 0.026 \\
    $\hat{\tau}_Y$ & 0.004 & 0.004 & 0.010 & 0.010 & 0.012 & 0.012 \\
   \hline
   \hline
    \end{tabular}%
      \caption{MCSD under different simulation designs and parameter settings. The true value of $(\theta, \tau_Y, \tau_X)=(0.2,0.2, 0.6)$. MAGIC$^*$ is the MAGIC estimator under the oracle scenario, and DMVMR$^*$ is DMVMR under the oracle scenario. The total number of SNPs is 100,000. Here $\pi_{{\mathtt{x}}\setminus {\mathtt{m}}}$ is the proportion of SNPs that are only associated with $X$,  $\pi_{{\mathtt{m}}\setminus {\mathtt{x}}}$ is the proportion of SNPs that are only associated with $M$,  $\pi_{{\mathtt{x}}\cap {\mathtt{m}}}$ is the proportion of SNPs that are  associated with both $X$ and $M$. }
  \label{tab: Variance for oracle scenario}%

\end{table}%

\section{Real data analysis}
\label{Sec:real-data-analysis}

Cardiovascular diseases (CVD) remain the leading global cause of death \citep{smith2012our}, with a high body-mass index (BMI) recognized as an important risk factor \citep{tsao2022heart}. However, the efficacy of current behavioral weight management interventions remains limited to the short term, and many weight loss medications, including recently approved ones, either lack long-lasting benefits or pose safety concerns \citep{tak2021long,myers2023effectiveness}. By contrast, there are many effective clinical and public health interventions available to control cholesterol levels, blood pressure, and fasting glucose levels \citep{finucane2011national}. This situation has sparked growing interest in identifying metabolic factors that mediate between high BMI and cardiovascular disease, with the ultimate goal of targeting these metabolic factors to reduce the risk of CVD. While several metabolic factors that mediate the adverse effects of high BMI on cardiovascular diseases development have been identified \citep{lu2015mediators}, most of these findings are from traditional observational studies, potentially suffering from unmeasured confounding issues \citep{richiardi2013mediation}. 

In this context, to identify metabolic factors that mediate the relationship between high BMI and cardiovascular diseases, we conduct mediation analysis with Mendelian randomization using GWAS summary data  from the IEU OpenGWAS project \citep{lyon2021variant} and MEGASTROKE consortium \citep{malik2018multiancestry}. The exposure we consider include BMI, indicator of obesity (BMI $\geq 30 \text{kg}/\text{m}^2$),  and waist-to-hip ratio (WHR). Our investigation focuses on two major cardiovascular diseases: coronary artery disease (CAD) and stroke. For potential mediators, we consider modifiable metabolic factors such as lipids (hypercholesterolaemia, low-density lipoprotein cholesterol (LDL), high-density lipoprotein cholesterol (HDL)), blood pressure (systolic blood pressure (SBP), diastolic blood pressure (DBP)), and blood glucose (fasting glucose), along with negative control factors (hair color before graying). Detailed data information is summarized in Supplementary Material Section 8.5. 

As a data pre-processing step, before conducting IV selection, we harmonize the exposure, mediator, and outcome GWAS data following the procedure detailed in Supplementary Material Section 8.1. Note that similar to the previous section, the IV selection procedure of MAGIC is different from the existing MR-based mediation analysis methods. Other than selecting relevant IVs, we also conduct clumping to remove the correlation between different IVs; see Supplementary Material Section 8.1 for implementation details.

\begin{table}[H]
  \centering
       \caption{Benjamini-Hochberg adjusted $p$-values for the mediation effects obtained from MAGIC and the two-step MR.}
\label{tab:total real data result}
    \begin{tabular}{ccccccc}
   \hline
   \hline
      \multicolumn{7}{c}{Metabolic Mediators} \\
      \hline
    \makecell{Mediators} & \makecell{Pure\\ hypercholes-\\-terolemia} &  LDL  & HDL   & SBP   &  DBP  &  \makecell{Fasting\\ Glu}  \\
     \hline
    \multicolumn{7}{c}{BMI$\rightarrow$ Mediator $\rightarrow$ CAD} \\
   MAGIC & 0.000 & 0.075 &0.000 &0.000& \textbf{0.000} & 0.783  \\
   Two-step MR & 0.008 & 0.056 & 0.000&0.004&  \textbf{0.064} & 0.079  \\
   
     \hline
    \multicolumn{7}{c}{WHR $\rightarrow$ Mediator $\rightarrow$ CAD} \\
 MAGIC & 0.050 & 0.494 & \textbf{0.050} &0.532 & \textbf{0.050} & 0.532\\
    Two-step MR & 0.024 &0.999& \textbf{0.106} & 0.198& \textbf{0.064} & 0.214  \\
    \hline
    \hline
    \end{tabular}%
\end{table}%

Given we have multiple mediators under consideration, we report Benjamini-Hochberg (BH) adjusted $p$-values \citep{benjamini1995controlling} in Table \ref{tab:total real data result}. There, we compare the results provided by MAGIC and the two-step MR. Our results showcase that MAGIC successfully identifies DBP as a significant mediator in two pathways: (i) BMI$\rightarrow$DBP$\rightarrow$CAD, and (ii) WHR$\rightarrow$DBP$\rightarrow$CAD, and identifies HDL as a significant mediator in the pathway WHR$\rightarrow$HDL$\rightarrow$CAD. Existing literature provided some supportive evidence of these pathways. For example, both DBP and SBP's role in mediating the risk of obesity to both CAD and stroke is well-established and the treatment of blood pressure has served as a main prevention and intervention therapy for CVD  \citep{zang2022intensive,mcmackin2007effect,ovbiagele2011level,macmahon1990blood}. 
It has also been found that HDL is a marker and a mediator of CAD by removing overloaded cholesterol from cells in the artery wall and exerting anti-inflammatory effects \citep{heinecke2009hdl}. In contrast, the two-step MR fails to identify these significant mediators, suggesting that MAGIC may have a higher detection power than the two-step MR approach. 

Additional results for other potential mediators including MVMR methods are provided in Section 8.2-8.4 in the Supplementary Material and are omitted in the main manuscript due to space limit. 

\section{Conclusion}

In this manuscript, we introduced a novel mediation analysis framework employing Mendelian randomization with summary data. Our framework efficiently integrates information from three GWAS with carefully crafted estimating equations, leading to accurate direct and mediation effect estimation with enhanced statistical efficiency. In addition, the proposed method is immune to the winner's/loser's curse, corrects the measurement error bias, and remains valid even when instrument selection is imperfect. We provided rigorous statistical guarantees, including a joint asymptotic normality characterization of the estimated direct and mediation effects. We further demonstrated the construction of valid standard errors. We also discussed the potential efficiency gains of our approach relative to the debiased multivariable Mendelian Randomization in an oracle setting. 

\section*{Acknowledgement}
This research was partially supported by the NSF Awards DMS-2239047 and DMS-2220537.

{\small
\onehalfspacing\bibliographystyle{jasa}
	\bibliography{reference}}
\end{document}